\documentclass[12pt,preprint]{aastex}
\usepackage{amssymb,amsmath,mathrsfs}
\usepackage{subfigure}

\shorttitle{\textsc{data-driven spectral modeling}}
\shortauthors{\textsc{tsalmantza \& hogg}}
\newcommand{\project}[1]{\textsl{#1}}
\newcommand{\sdss}{\project{SDSS}}
\newcommand{\SDSS}{\sdss}
\newcommand{\documentname}{\textsl{Article}}
\newcommand{\sectionname}{Section}
\newcommand{\equationname}{equation}
\newcommand{\etal}{\textit{et~al.}}
\newcommand{\tv}[1]{\boldsymbol{#1}}
\newcommand{\inverse}[1]{{#1}^{-1}}

\begin{document}\sloppy


\title{A data-driven model for spectra:\\
       Finding double redshifts in the \project{Sloan Digital Sky Survey}}
\author{P.~Tsalmantza\altaffilmark{1} and David~W.~Hogg\altaffilmark{1,2}}
\email{vivitsal@mpia.de}
\altaffiltext{1}{Max-Planck-Institut f\"ur Astronomie, K\"onigstuhl 17, 69117 Heidelberg, Germany}
\altaffiltext{2}{Center~for~Cosmology~and~Particle~Physics, Department~of~Physics, New~York~University, 4~Washington~Place, New~York, NY 10003, USA}

\begin{abstract}
We present a data-driven method---heteroscedastic matrix
factorization, a kind of probabilistic factor analysis---for modeling
or performing dimensionality reduction on observed spectra or other
high-dimensional data with known but non-uniform observational
uncertainties.  The method uses an iterative inverse-variance-weighted
least-squares minimization procedure to generate a best set of basis
functions.  The method is similar to principal components analysis,
but with the substantial advantage that it uses measurement
uncertainties in a responsible way and accounts naturally for poorly
measured and missing data; it models the variance in the
noise-deconvolved data space.  A regularization can be applied, in the
form of a smoothness prior (inspired by Gaussian processes) or a
non-negative constraint, without making the method prohibitively slow.
Because the method optimizes a justified scalar (related to the
likelihood), the basis provides a better fit to the data in a
probabilistic sense than any PCA basis.  We test the method on
\SDSS\ spectra, concentrating on spectra known to contain two redshift
components: These are spectra of gravitational lens candidates and
massive black-hole binaries. We apply a hypothesis test to
compare one-redshift and two-redshift models for these spectra,
utilizing the data-driven model trained on a random subset of all
\SDSS\ spectra.  This test confirms 129 of the 131 lens candidates in
our sample and all of the known binary candidates, and turns up very
few false positives.
\end{abstract}


\section{Introduction}\label{sec:introduction}
Data-driven models are necessary for many applications: It is rare
that theoretical models are specific enough, accurate enough, or rich
enough to generate all of the features of a data set.  Common examples 
in astronomy come in the study and analysis of spectra.  Redshift
determination (\citealt{budavari}; \citealt{glazebrook}),
emission-line measurements and properties of objects (\citealt{allen}; 
\citealt{borosona}; \citealt{borosonc}; \citealt{wild}), 
decomposition into different stellar populations (\citealt{chen}; 
\citealt{ferreras}; \citealt{nolan}) and
classification of sources (\citealt{borosona}; \citealt{connolly};
\citealt{francis}; \citealt{suzuki}; \citealt{yip}; \citealt{yipb}) can all in
principle be performed with theory-based spectral models, but usually these 
models are not accurate or detailed enough to make
measurements as precise as contemporary data permit.  In addition to
the quantitative deficiencies of theoretical models (the fact that
they rarely can explain the data at the precision of the
observations), theoretical models contain qualitative
uncertainties---uncertainties about model assumptions and
computational approximations---that inevitably propagate into results.
For these reasons, the highest-performing redshift determination
systems, for example, use data-driven models that involve principal
components analysis (PCA) or similar approaches (e.g., \citealt{sdssdr7}).

Data-driven models like PCA are excellent for describing the range of
the data---the subspace of the (usually enormous) data space in which
real data examples live.  For this reason, data-driven models are
excellent for finding \emph{outliers}, objects that are unusual in the
data set.  This application is not well explored in astrophysics,
but one of the motivations of the present investigation is to explore
the use of probabilistically justified data-driven models for outlier
detection.

Similarly, data-driven models are often used for classification.  When
a data-driven model produces ``eigenspectra'' or clusters or
equivalent, it is tempting to see these model properties as defining
classes in the data.  This gets into the area of unsupervised
classification, which is beyond the scope of the present work.
We will just comment here that when a
generic data-driven model is being used without theoretical
justification (i.e. the probabilistic objective function used by the 
method has no straightforward relation with the underlying physical properties 
of the sources), it is often a mistake to interpret the internals of
that model physically, however tempting that may be. This is a 
common problem for all unsupervised data-driven methods (including the 
method presented here, PCA, Non-negative matrix factorization etc.).

The standard data-driven models used for spectroscopic astronomical
data are the highest-ranked principal components (from PCA or
equivalent). PCA has a few advantages and a number of drawbacks.  The
advantages are that it is convex---there is only one optimum of the PCA
objective function---and that it is
entirely data-driven: The construction of the
PCA requires no theoretical or external knowledge about the spectra
being modeled; it is a dimensionality reduction in the space of the
observed spectra.

There are many drawbacks to PCA but the most important is that the PCA
returns the principal directions---the eigenvectors with maximum
eigenvalues---of the \emph{variance tensor} of the data; this variance
tensor has contributions from intrinsic variation among spectra, and
contributions from observational noise.  That is, a direction in
spectrum space can enter into the top principal components because it
is a direction of great astrophysical variation, or because there is a
lot of noise in the observations along that direction, or both.  PCA
is \emph{agnostic} about the source of the variance, while astronomers are
\emph{not}; astronomers have knowledge of the amount of variance caused by 
observational noise. By using a method that is taking advantage of this 
knowledge we can obtain more realistic results that can reproduce better the properties 
of the data.

Other drawbacks to PCA include the following: It treats the data as
having been drawn from a linear subspace of the full spectral
space. This assumption is unlikely to be true in any application. 
As an example, in the problem of modeling spectra that we are 
examining in the present study, the relation between the spectral flux 
and the physical parameters is highly non-linear \citep{vanderplas}.  It
also has trouble separating the spectral variation that comes from
amplitude changes (overall flux or luminosity changes) as distinct
from variations that come from shape changes in the spectra.  Various
hacks have been employed to deal with this, but many of them make the
linear subspace assumption even less valid than it was \textit{a
  priori}. Finally, PCA has no idea about prior information; it is
just as happy creating components with negative amplitudes as positive
amplitudes and the linear subspace therefore contains many quadrants,
in general, that represent spectra with completely unphysical
properties (such as negative emission lines and the like).

In this \documentname, we introduce a new---or at least new to
astrophysics---data-driven technique, heteroscedastic matrix
factorization (HMF), for modeling observed spectra that overcomes some
(though not all) of the problems with PCA.  The principal advantage of
HMF over PCA is that the method optimizes not squared error, but
rather the justified probabilistic objective of chi-squared.  This
leads to important differences with PCA: \textsl{(1)}~HMF builds a
model of the variance in the data set \emph{not} introduced by
observational noise.  That is, it returns a model of the
noise-deconvolved spectral space, which is the space of interest to
the scientific investigator.  \textsl{(2)}~HMF deals absolutely
naturally with the completely generic problems of missing and badly
measured data; no ``patching'' of the method or the input data is
required to use HMF.

For dimensionality reduction or data-driven modeling of spectra (or
any other data), the investigator has an enormous number of options.
There are PCA, K-means, Independent Component Analysis (ICA), 
and factor analysis to name just a few (\citealt{rz}).
HMF is an example of a probabilistic factor analysis;
it is probabilistic because it optimizes an objective that
(unlike the objective for PCA or K-means or standard ICA) is
justified in terms of a likelihood; it is a factor analysis because it
reduces the dimensionality of the data matrix to a product of two
low-rank factors.  It is our view that if the output of a
dimensionality reduction or data-driven model is going to be used in
probabilistic inference, it ought to emerge from a probabilistically
justified method.

Because these ideas are so wide-spread there is a lot of prior art;
for a tour of the matrix factorization or dimensionality reduction
methods and their relationships, excellent reviews exist
(\citealt{rz}).
Similar methods to the HMF presented here have been
published in the computer vision literature, in machine 
learning (\citealt{wilsona, wilsonb}), in bioinformatics \citep{bio} and in
the chemistry literature (as ``maximum-likelihood PCA'';
\citealt{mlpca}, although without the particular regularizations we
propose).  The internal model of the \texttt{kcorrect} software
\citep{blanton} is a version of HMF with non-negativity constraints
applied; the optimization methods we use below for non-negative HMF
are adapted from that source.

While it is relatively novel (in astrophysics) to be replacing the PCA
objective function with one that is probabilistically justified, it is
not new to be concerned about the missing-data problems with PCA (\citealt{connollya, budavarirobust}).
Along these lines, various investigators have developed ``patching''
techniques to sensibly replace missing data (for example,
\citealt{eishogg, wild, budavarirobust, borosona}); these methods are
heuristic and bias the results relative to any probabilistic treatment
in unknown ways.

Though better than PCA for the reasons given above, HMF does not
directly address its linearity and prior-PDF problems.  However,
because the objective function has a direct likelihood interpretation,
it becomes possible to incorporate the output of HMF into a Bayesian
inference with properly informative prior information.

At the beginning, we mentioned outlier identification.  Some of the
most important outliers found in the \sdss\ data are double-redshift
objects.  Indeed, the set of luminous red galaxies that show evidence
for a second (higher) redshift in their spectra include as a subset a
significant fraction of all known gravitational lenses \citep{bolton}.
Other valuable double-redshift sources include merging galaxies,
binary stars, and binary quasars.  In the latter category, there are
very few known examples where the best explanation for a spectrum is a
bound pair of massive black holes (\citealt{komossa};
\citealt{shields}: \citealt{boroson}; \citealt{decarli}; \citealt{barrows11}; 
\citealt{eracleous}; \citealt{tsalmantza}).  In most of the cases, the best
candidates  of gravitational lenses and binary black holes have been 
found with heuristic searches.  These searches
involve (in the case of gravitational lenses) looking for isolated
emission lines at the second redshift, or (in the case of black-hole
binaries) visual inspection of double or shifted broad emission lines.
In both cases, very high quality data-driven models ought to make the
searches more sensitive and more complete.

In what follows, we introduce the HMF data-driven model and methods
for implementing it.  We test the model and methods on \sdss\ spectra.
We assess the value of the HMF model by asking whether it confirms the
known double-redshift objects in the \sdss\ spectroscopic data set.

\section{Spectral model}\label{sec:model}
Each of the $N$ observed spectra $i$ can be thought of as an ordered
list or column vector $\vec{f}_i$ of $M$ flux density (energy per area
per time per wavelength) measurements $f_{ij}$ on a grid of $M$
observer-frame wavelengths $\lambda^{\mathrm{obs}}_j$:
\begin{equation}
\vec{f}_i
\equiv \left[\begin{array}{c} f_{i1} \\
                              f_{i2} \\
                              \cdots \\
                              f_{iM} \end{array}\right]
\equiv \left[\begin{array}{c} f_{\lambda,i}(\lambda^{\mathrm{obs}}_1) \\
                              f_{\lambda,i}(\lambda^{\mathrm{obs}}_2) \\
                                                \cdots \\
                              f_{\lambda,i}(\lambda^{\mathrm{obs}}_M) \end{array}\right]
\quad ,
\end{equation}
Associated with each measurement $f_{ij}$ is an uncertainty variance
$\sigma_{ij}$ and we will assume in what follows that these
uncertainty variances are well measured and that the uncertainties are
essentially Gaussian.  We will assume that off-diagonal terms
(covariances) in the uncertainty variance tensor are small, or that
the uncertainty variance tensor (covariance matrix) $\tv{C}_i$ is
approximately
\begin{equation}
\tv{C}_i =
 \left[\begin{array}{cccc} \sigma_{i1}^2 & 0 & & 0 \\
                           0 & \sigma_{i2}^2 & & 0 \\
                           & & \cdots & \\
                           0 & 0 & & \sigma_{iM}^2 \end{array}\right]
\quad .
\end{equation}

We want to model the spectrum of each object $i$ with a sum of $K$
linear components:
\begin{equation}\label{eq:model}
f_{ij} = f_{\lambda,i}(\lambda_j) = \sum_{k=1}^{K} a_{ik}\,g_k(\lambda_j) + e_{ij}
\quad ,
\end{equation}
where the modeling is done implicitly in the object rest frame, the
$a_{ik}$ are coefficients, the $g_k(\lambda)$ are basis spectra, and
the $e_{ij}$ represents the individual noise in pixel $j$ of spectrum
$i$.  The noise element $e_{ij}$ is assumed to be drawn from a
Gaussian of zero mean and variance $\sigma_{ij}^2$.  The
dimensionality $K$ is an investigator-set model complexity parameter,
the objective setting of which is discussed briefly below.  Given
basis spectra $g_k(\lambda)$, the best set of coefficients for any
observed spectrum---under the assumption of known, Gaussian
uncertainties---can be found by weighted least-square fitting.  The
challenge is to find the best set of basis spectra.

Often in astronomy, this basis has been found by principal components
analysis (or equivalent) and then selection of the largest-variance
components or largest-eigenvalue eigenvectors.  However, as we
emphasize in \sectionname~\ref{sec:introduction}, this use of PCA
naturally locates the $K$-dimensional linear basis that minimizes the
mean-squared error in the space in which all pixels of all spectra are
treated equally: They are weighted equally in the analysis, and
residuals in them are minimized by the PCA with equal aggression. This
is an inappropriate approach in the real situation in which different
data points come with very different uncertainty variances, and it is
absolutely inapplicable when there are missing data---as there always
are in real data sets.

For these reasons, we seek to find the basis set that optimizes a
justified scalar objective, one that is consistent with the individual
spectral pixel uncertainty variances and with the fact that there are
missing data.  When uncertainties are close to Gaussian with known
variances, the logarithm of the likelihood is proportional to
chi-squared, so we seek to find the basis functions and coefficients
that minimize a total chi-squared:
\begin{eqnarray}\label{eq:chi-squared}\displaystyle
\chi^2 & = & X - 2\,\ln p(d|m) \nonumber\\
\chi^2 & = & \sum_{i=1}^N \sum_{j=1}^M
             \frac{\left[f_{ij}-\sum_{k=1}^K a_{ik}
                      \,g_k(\lambda_j/[1+z_i])\right]^2}
{\sigma^2_{ij}}
\quad ,
\end{eqnarray}
where $X$ is a constant, $p(d|m)$ represents the likelihood function
(probability of all the data given the model), and we have implicitly
assumed that each spectrum $i$ under consideration at this stage is
well explained by having all its flux come from a single object at a
known redshift $z_i$. The model contains the $N\,K$ coefficients
$a_{ik}$ and the $K$ functions $g_k(\lambda)$.

The model of \equationname~(\ref{eq:model}) is a \emph{matrix
  factorization} in the sense that if we think of the full set of data
$f_{ij}$ as comprising a large rectangular matrix, the coefficients
and basis functions provide a low-rank outer-product approximation to
that matrix.  The scalar objective $\chi^2$ is \emph{heteroscedastic}
in that it takes account of the fact that each matrix element has a
noise contribution with a different expected variance.

Roughly speaking, we seek to find the coefficients $a_{ik}$ and basis
functions $g_k(\lambda)$ that globally minimize the scalar $\chi^2$.
Precisely speaking, we make two adjustments to this goal. The first is
that we can't demand global optimization; this problem is not convex.
Indeed, there are enormous numbers of local minima, in both the
trivial sense that there are exact degeneracies (swap two basis
functions and their corresponding coefficients, or re-scale a basis
function and the corresponding coefficients, and so on) and in the
non-trivial sense that there are multiple qualitatively different
optima. All that our methods (described in detail below) guarantee is
that we have, at fixed coefficients $a_{ik}$ the globally optimal
basis functions $g_k(\lambda)$ and that we have, at fixed basis
functions $g_k(\lambda)$ the globally optimal coefficients $a_{ik}$.

\paragraph{Regularization:}
The second adjustment is that we sometimes choose to impose a
regularization to improve the performance or realism of the model.
The first kind of regularization we can impose is a smoothness prior
that improves performance at (rest-frame) wavelengths at which we have
very few data. In practice, we implement this prior by constructing
the basis functions $g_k(\lambda)$ on a grid of $M$ rest-frame wavelengths
$\lambda_j$ and penalizing quadratically large pixel-to-pixel
variations. That is, we optimize not the pure $\chi^2$ above but a
modified scalar $\chi_{\epsilon}^2$
\begin{equation}\label{eq:smoothness}
\chi_{\epsilon}^2 \equiv \chi^2
 + \epsilon\,\sum_{k=1}^K \sum_{j=2}^{M}
 \left[g_k(\lambda_{j})-g_k(\lambda_{j-1})\right]^2
\quad ,
\end{equation}
where $\chi^2$ is defined in \equationname~(\ref{eq:chi-squared}),
$\epsilon$ is a scalar that sets the strength of the smoothing.  The
investigator can set $\epsilon$ to tune the smoothness of the model;
setting this parameter objectively is discussed briefly below.
Optimization of this scalar $\chi_{\epsilon}^2$ is equivalent to
optimization of the posterior probability distribution with a Gaussian
prior applied to the pixel-to-pixel differences \citep[for
  example,][]{smoothness}; it is similar to what is done with Gaussian
Processes \citep[for example,][]{Rasmussen06a}.

The second kind of regularization we can impose is non-negativity.
That is, we can optimize \equationname~(\ref{eq:chi-squared}) but subject
to the constraint that all basis functions and all coefficients are
non-negative:
\begin{eqnarray}\label{eq:non-negative}\displaystyle
a_{ik} & \geq & 0 \quad\mbox{for all $i$, $k$} \nonumber\\
g_{k}(\lambda) & \geq & 0 \quad\mbox{for all $k$, $\lambda$}
\end{eqnarray}
This can lead to solutions that are much more physically meaningful,
especially when the objects of study are astronomical spectra of
galaxies (and particularly if the galaxies are composed of components
that are themselves optically thin).

\paragraph{Model complexity:}
The model has $[N\,K + M\,K]$ free parameters, where $N$ is the number
of $M$-dimensional data points (set by the size of the data set), $M$
is the number of wavelengths (set by the size of each data point), and
$K$ is the number of components permitted, or the dimensionality of
the model space.  The dimension $K$ has to be chosen either
arbitrarily by the investigator, or else by a process of model
selection.

Standard frequentist model selection methods (for example, the AIC
\citep{aic} or the BIC \citep{bic}) 
compare the objective $\chi^2$ of
\equationname~(\ref{eq:chi-squared}) to some re-scaling of the
\emph{number of degrees of freedom}, which is the number of independent
data measurements ($[N\,M]$ in this case) minus the number of free
parameters ($[N\,K + M\,K]$ in this case).  These model-selection
methods are simple to implement but have many problems
including---but not limited to---the following: They are only
justified by certain un-stated \emph{utility} assumptions; model
selection requires a utility, and these leave that utility unstated.
They are only justified in the case of pure linear fitting, and this
model is not linear; it is bi-linear.  The number of measurements is
not a well-defined quantity; for example it is not clear what to do
about measurements $f_{ij}$ for which the inverse variance
$1/\sigma^2_{ij}$ is very small, or zero.  In practice, also, we find
that the AIC and BIC also tend to prefer very high $K$, even in
situations where the data are clearly well-described by a model with
low $K$.  This may have to do with the assumption of Gaussianity; the
AIC and BIC are strongly pulled by outliers in the data.

The standard Bayesian model selection involves performing an integral
of the likelihood (which is an exponential of $\chi^2$) over the prior
to obtain the \emph{evidence} for each model.  This method also has
several problems including the following: A proper prior must be
specified that is a function of the parameters \emph{and} the
dimensionality $K$; this is not only difficult, arbitrary
specification can lead to spurious results.  That is, the prior needs
to be not just proper but in fact \emph{justified} or properly
informative.  Evidence calculations require (in principle) integration
over the entire permitted model space, which is enormous in this case;
these integrals are rarely possible.  Furthermore, even done
correctly, the evidence integral gives the relative probabilities of
the models, but that does not suffice for model selection, which
requires a specified utility.

For all these reasons, we recommend leave-one-out cross-validation or
some similar kind of train-and-test framework.  In this technique, the
model is fit to all but a ``left out'' subset of the data, and then
the best-fit model is used to predict or model the left-out subset.
If $K$ is too small, the model doesn't have enough freedom to fit even
the training set well, and if $K$ is too large, the model over-fits
the training set and does worse on the test set.  That is, we
recommend choosing the dimensionality $K$ where the model does the
best at predicting new or left-out data.  This is a good,
scientifically motivated utility.  Also, typically, running $L$ trials
of leave-one-out usually takes less than or of order $L$ times as long
as the original optimization of the model, so it is not expensive.
Finally, we find in practice that the cross-validation-optimal model
complexity seems intuitively reasonable.

When the smoothness regularization of
\equationname~(\ref{eq:smoothness}) is turned on, the smoothness
parameter $\epsilon$ is also a model-complexity parameter; if
$\epsilon$ is set near zero the model has the full bi-linear freedom;
if $\epsilon$ is set large, the model cannot easily explore ``spiky''
parts of the parameter space.  This continuous model-complexity
parameter doesn't fit into the AIC or BIC framework (which needs a
clearly stated \emph{number} of degrees of freedom to operate); this
parameter must be set by a Bayesian-evidence or cross-validation
methods.  Once again, we advocate cross validation.

All that said, there is no need to set the model-complexity parameters
$K$ and $\epsilon$ by any objective model-selection test.  For many
purposes, there will be clear limits to what is possible from hardware
and computer-time constraints, or ball-park estimates of what makes
sense that are good enough.  Indeed, our cross-validation tests
suggest that in most scientific situations, the quality of the model
is a very slow function of $K$ and $\epsilon$.  Heuristic (that is,
by-eye or by-hand) setting of
$K$ and $\epsilon$ is therefore legitimate in many or most situations, as long as there is not strong dependence of the results on these model-complexity parameters near the chosen settings.

\section{Optimization}\label{sec:optimization}

The HMF spectral model is defined by the coefficients $a_{ik}$ and basis
functions $g_k(\lambda)$ in \equationname~(\ref{eq:model}) and the
objective function in \equationname~(\ref{eq:chi-squared}), with possible
regularization for smoothness given in
\equationname~(\ref{eq:smoothness}) or non-negativity given in
\equationname~(\ref{eq:non-negative}).  Optimization of the model is an
engineering problem that is in principle independent of the arguments
for the model freedom or the objective function.  In practice, there
is a natural optimization technique for bi-linear models like this
one, which we describe here.  As mentioned above, the problem is not
convex, so the \emph{initialization} for the optimization matters.  We
leave discussion of the initialization to
\sectionname~\ref{sec:initialization}.

In principle there are many parameterizations for the basis functions
$g_k(\lambda)$.  For any implementation of the optimization methods
given here, it makes sense for the parameterization of these basis
functions to be \emph{linear}.  For our specific purposes here, where
redshifting is an important task, it makes sense for the the basis
functions to be evaluations of the basis spectra on a wavelength grid
that is linearly spaced in log-wavelength.  That is, we set
coefficients $g_{kj}$ defined by
\begin{eqnarray}\displaystyle
g_{kj} & \equiv & g_k(\lambda_j) \nonumber\\
\ln\lambda_j & = & \ln(\lambda_0) + j\,\Delta
\quad ,
\end{eqnarray}
where $\lambda_0$ is the minimum wavelength under consideration, and
$\Delta$ is a logarithmic wavelength interval.  The basis spectrum
$g_k(\lambda)$ is defined to be the cubic-spline interpolation in
logarithmic wavelength $\ln\lambda$ between the control points
$g_{kj}$ at wavelengths $\lambda_j$ to the wavelength of interest
$\lambda$.  In what follows, much of the fitting is in rest-frame
wavelengths, given observations in observed-frame wavelengths.  The
logarithmic wavelength grid and cubic spline interpolation makes
redshift transformations simple and fast.  It is also the case that the
\sdss\ spectra used below are extracted on a logarithmic cubic-spline
basis of this kind.  In detail, we will have to put all the input data
and basis spectra on the same wavelength grid.

From any initialization, we can descend to a local minimum using
iterated least squares.  To begin, consider the unregularized case,
that is, when the objective function is $\chi^2$ of
\equationname~(\ref{eq:chi-squared}) and no constraints are applied to
the parameters.  In this case, we optimize as follows: In each step,
we fix the $g_{kj}$ and find the optimal $a_{ik}$ by weighted least
squares, and then hold the $a_{ik}$ fixed and find the optimal
$g_{kj}$ by weighted least squares. Each iteration step is guaranteed
to reduce the total $\chi^2$ and converges (in practice in the tests
below) in ten or so iterations. We should note that by convergence 
here we mean that the iterated algorithm reaches a stationary point.

There are $[N\,K]$ coefficients $a_{ik}$ and $[K\,M]$ parameters
$g_{kj}$ in the basis spectra.  For most data sets, least-square fits
with this many parameters are impossible with na\"ive algorithms.
However, the relevant matrices are extremely sparse, in the sense that
each parameter only affects a small number of data points $f_{ij}$.
Any high-quality sparse-matrix linear algebra system can efficiently
solve the full weighted least-square minimization problem for all the
$a_{ik}$ in one shot (what we will call the ``a-step''), and then the
full problem for all the $g_{kj}$ (the ``g-step'') in a second shot.
Then iteration can proceed to convergence.  However, the sparseness is
very simple, so in the absence of a sparse-matrix linear algebra
solver, the a-step and g-step can be split into iterated blocks.  The
block-diagonal a-step involves an iteration over objects $i$, and the
block-diagonal g-step involves an iteration over wavelength indices
$j$.

The \emph{block-diagonal a-step} is, for each $i$,
\begin{eqnarray}\label{eq:astep}\displaystyle
\tv{A}_i & \gets & \inverse{\tv{G}_i}\cdot\tv{F}_i \nonumber\\
\left[\tv{A}_i\right]_k & \equiv & a_{ik} \nonumber\\
\left[\tv{G}_i\right]_{kk'} & \equiv & \sum_{j=1}^{M} \frac{g_{kj}\,g_{k'j}}{\sigma_{ij}^2} \nonumber\\
\left[\tv{F}_i\right]_k     & \equiv & \sum_{j=1}^{M} \frac{g_{kj}\,f_{i j}}{\sigma_{ij}^2}
\quad ,
\end{eqnarray}
where at fixed $i$, $\tv{A}_i$ is a vector of the $K$ coefficients
$a_{ik}$, $\tv{G}_i$ is a $K\times K$ matrix to be inverted, and
$\tv{F}_i$ is a $K$-vector.  This is just weighted linear least
squares at each index $i$, fixing the $g_{kj}$.

The \emph{block-diagonal g-step} is, for each $j$,
\begin{eqnarray}\label{eq:gstep}\displaystyle
\tv{G}_j & \gets & \inverse{\tv{A}_j}\cdot\tv{F}_j \nonumber\\
\left[\tv{G}_j\right]_k & \equiv & g_{kj} \nonumber\\
\left[\tv{A}_j\right]_{kk'} & \equiv & \sum_{i=1}^{N} \frac{a_{ik}\,a_{ik'}}{\sigma_{ij}^2} \nonumber\\
\left[\tv{F}_j\right]_k     & \equiv & \sum_{i=1}^{N} \frac{a_{ik}\,f_{i j}}{\sigma_{ij}^2}
\quad ,
\end{eqnarray}
where at fixed $j$, $\tv{G}_j$ is a vector of the $K$ spectral element
values $g_{kj}$, $\tv{A}_j$ is a $K\times K$ matrix to be inverted, and
$\tv{F}_j$ is a $K$-vector.  Note the parallelism with the a-step
\equationname~(\ref{eq:astep}).

Each of these iterated steps, in turn, optimizes one part of the
problem leaving the other fixed; each step therefore reduces the
scalar objective $\chi^2$ of \equationname~(\ref{eq:chi-squared}); the
pair of iterated steps can be iterated until the system reaches a
minimum.  These steps must be modified when smoothness
(\equationname~\ref{eq:smoothness}) or non-negativity
(\equationname~\ref{eq:non-negative}) regularizations are applied, as
we discuss next.

For any $K$-dimensional linear subspace, there are many choices for
the components $g_k$: The components can be reordered, multiplied by
scalars, or replaced with linear combinations of themselves.  We don't
try to break all of these degeneracies, but we do enforce the $K$
constraints that the $\vec{g}_k\cdot\vec{g}_k=1$ or $M$ or equivalent.
Additionally, in a way analogous to the PCA, we orthogonalize the
system of the basis functions and reorder the components according to
the variance they include by diagonalize the squared matrix of their
coefficients when fit to the data.  More specifically we use the
$K\times K$ matrix U that includes the eigenvectors of the $A^TA$
matrix to define the new components and coefficients as follows:

\begin{eqnarray}\label{eq:ordering}
a_{ik}=\sum_{k'=1}^{K}a_{ik'}u_{k'k} \nonumber\\
g_{kj}=\sum_{k'=1}^{K}u_{kk'}g_{k'j}
\end{eqnarray}

\paragraph{Optimization with smoothness prior:}

The smoothness regularization---the move to the $\chi^2_{\epsilon}$
scalar objective of \equationname~(\ref{eq:smoothness})---reduces the
sparseness of the linear system.  For this reason, the simple blocking
of the problem in {\equationname}s~(\ref{eq:astep}) and
(\ref{eq:gstep}) is no longer possible as an exact solution.  However,
it is possible to approximate an exact solution by modifying the
block-diagonal g-step to:
\begin{eqnarray}\label{eq:gstep_smooth}\displaystyle
\tv{G}_j & \gets & \inverse{\tv{A}_j}\cdot\tv{F}_j \nonumber\\
\left[\tv{G}_j\right]_k & \equiv & g_{kj} \nonumber\\
\left[\tv{A}_j\right]_{kk'} & \equiv & \sum_{i=1}^{N} \frac{a_{ik}\,a_{ik'}}{\sigma_{ij}^2} + 2\,\epsilon\,\delta_{kk'} \nonumber\\
\left[\tv{F}_j\right]_k     & \equiv & \sum_{i=1}^{N} \frac{a_{ik}\,f_{i j}}{\sigma_{ij}^2} + \epsilon\,[g_{k[j-1]} + g_{k[j+1]}]
\quad ,
\end{eqnarray}
where $\delta_{kk'}$ is the $K\times K$ identity matrix, and where
small changes need to be made at $j=1$ and $j=M$:
\begin{eqnarray}\label{eq:gstep_tweaks}\displaystyle
\left[\tv{A}_1\right]_{kk'} & \equiv & \sum_{i=1}^{N} \frac{a_{ik}\,a_{ik'}}{\sigma_{i1}^2} + \epsilon\,\delta_{kk'} \nonumber\\
\left[\tv{F}_1\right]_k     & \equiv & \sum_{i=1}^{N} \frac{a_{ik}\,f_{i 1}}{\sigma_{i1}^2} + \epsilon\,g_{k2} \nonumber\\
\left[\tv{A}_M\right]_{kk'} & \equiv & \sum_{i=1}^{N} \frac{a_{ik}\,a_{ik'}}{\sigma_{iM}^2} + \epsilon\,\delta_{kk'} \nonumber\\
\left[\tv{F}_M\right]_k     & \equiv & \sum_{i=1}^{N} \frac{a_{ik}\,f_{i M}}{\sigma_{iM}^2} + \epsilon\,g_{k[M-1]}
\quad .
\end{eqnarray}
This new g-step is justified by seeing the neighboring $g_{kj}$ pixels
as ``data'' that constrain the model with inverse variance $\epsilon$.
In case it isn't obvious, in the above g-step, the $g_{k[j-1]}$ and
$g_{k[j+1]}$ used in the smoothness term are those computed in the
previous iteration.

Because the smoothness constraint makes the system less sparse, it
might become sensible to use conjugate gradient method \citep[for
  example,][]{shewchuk}, which permits optimization of least-squares
problems without explicit matrix decompositions or inversions.  When
using conjugate gradient descent we update the coefficients based with
a new \emph{conjugate-gradient a-step}:
\begin{eqnarray}\label{eq:astep_conj}\displaystyle
a_{ik} & \gets & a_{ik}+\alpha\,r_{ik} \quad \mbox{where} \nonumber\\
r_{ik} & = & \sum_{j=1}^{M}g_{jk}\,\frac{f_{ij}-\sum_{k=1}^{K}a_{ik}\,g_{kj}}{\sigma^2_{ij}} \nonumber\\
\alpha & = & \frac{\sum_{i=1}^{N}\sum_{k=1}^{K}r^2_{ik}}{\sum_{i=1}^{N}\sum_{k=1}^{K}r_{ik}\,R_{ik}} \nonumber\\
R_{ik} & = & \sum_{j=1}^{M}g_{jk}\,\frac{r_{ik}\,g_{kj}}{\sigma^2_{ij}}
\quad ,
\end{eqnarray}
and a new \emph{conjugate-gradient g-step}:
\begin{eqnarray}\label{eq:gstep_conj}\displaystyle
g_{kj} & \gets & g_{kj}+\beta\,q_{kj} \quad \mbox{where} \nonumber\\
q_{kj} & = & \sum_{i=1}^{N}a_{ik}\,\frac{f_{ij}-\sum_{k=1}^{K}a_{ik}\,g_{kj}}{\sigma^2_{ij}}+\epsilon\,[g_{k(j+1)}-g_{kj}]+\epsilon\,[g_{k(j-1)}-g_{kj}] \nonumber\\
\beta & = & \frac{\sum_{k=1}^{K}\sum_{j=1}^{M}q^2_{kj}}{\sum_{k=1}^{K}\sum_{j=1}^{M}q_{kj}\,Q_{kj}} \nonumber\\
Q_{kj} & = & \sum_{i=1}^{N}a_{ik}\,\frac{a_{ik}\,q_{kj}}{\sigma^2_{ij}}-\epsilon\,[q_{k(j+1)}-q_{kj}]-\epsilon\,[q_{k(j-1)}-q_{kj}]
\quad .
\end{eqnarray}
In our limited tests in the \project{R} language, we found that
conjugate gradient was much faster per iteration than the
block-diagonal method but required many more iterations to converge to
the same precision.  We therefore used the block-diagonal a-step and
g-step in everything that follows with the smoothness regularization.
But the relative performance of any real system running the one-shot
optmization with a sparse matrix representation, the block-diagonal
versions, or the conjugate-gradient version will in principle depend
on the numerical linear algebra system under use, the quantitative
properties of the data and its associated noise variances, and the
magnitudes of $N$, $M$, and $K$.

\paragraph{Optimization with non-negative constraints:}
If the non-negative constraint of
\equationname~(\ref{eq:non-negative}) is applied, normal weighted
least-squares techniques cannot be used; these know nothing about
constraints.  Fortunately, there are straightforward algorithms for
quadratic programming with linear constraints.  Indeed, as we
mentioned in \sectionname~\ref{sec:introduction}, the non-negative HMF
model has been used previously \citep{blanton}.  Optimization in this
case can proceed from all-positive initialization (to be discussed in
\sectionname~\ref{sec:initialization}) by purely multiplicative
updates.

The \emph{non-negative a-step} is:
\begin{equation}\label{eq:astep_nonneg}
a_{ik} \gets a_{ik}\,\left[\sum_{j=1}^{M}\frac{1}{\sigma^2_{ij}}\,f_{ij}\,g_{kj}\right]\left[\sum_{n=1}^{K}\sum_{j=1}^{M}\frac{1}{\sigma^2_{ij}}\,a_{in}\,g_{nj}\,g_{kj}\right]^{-1}
\quad ,
\end{equation}
and the \emph{non-negative g-step} is
\begin{equation}\label{eq:gstep_nonneg}
g_{kj} \gets g_{kj}\,\left[\sum_{i=1}^{N}\frac{1}{\sigma^2_{ij}}\,f_{ij}\,a_{ik}\right]\left[\sum_{n=1}^{K}\sum_{i=1}^{N}\frac{1}{\sigma^2_{ij}}\,a_{ik}\,a_{in}\,g_{nj}\right]^{-1}
\quad ,
\end{equation}
where in both cases the ``inverse'' is just a scalar inverse of each
term, not a matrix inverse of any kind.  These updates (for what might
be called non-negative HMF or NNHMF) were first written down in the
context of the $K$~correction system $\textsl{k-correct}$
\citep{blanton}.

As may be obvious from those equations, the presence of negative
fluxes $f_{ij}$ in the (noisy) observed spectra can lead to negative
solutions for the estimated components and coefficients. For this
reason, before we apply the method to the observed training set, all
the negative fluxes and their corresponding errors are set to zero. In
this way the data of those pixels is not taken into account when we
use the non-negative a-step and g-step.  Strictly speaking, this step
of removing negative fluxes is unjustified, but in most cases very few
pixels are affected, and the alternatives are substantially harder to
implement.

The non-negative a-step and g-step are very fast, permitting large
numbers of iterations ($\approx 10^3$), but that is good, because
convergence is generally slow.  To decrease the number of optimization
iterations, we find that it is beneficial to iterate the non-negative
a-step of \equationname~(\ref{eq:astep_nonneg}) many times ($\approx
10^2$) during the initialization while keeping the initial basis
$g_{kj}$ fixed, in order to start at a good set of initial
coefficients $a_{ik}$.

In the standard method (no regularization), the basis spectra define
an unconstrained linear subspace in which the spectra live; for this
reason, any (non-degenerate) linear combination of the basis spectra
constitute an equivalent basis.  When the non-negative constraint is
applied, this is no longer true; rotations, coadditions, or shears in
the basis-spectrum space will in general break non-negativity.  For
this reason, the basis spectra cannot be orthogonalized in any sensible 
way; at best they can be re-scaled to ensure
\begin{equation}\label{eq:nnnormalization}
\sum_{j=1}^{M}g_{kj}^2 = 1
\quad ,
\end{equation}
and ordered by decreasing variance such that
\begin{equation}\label{eq:nnordering}
\sum_{i=1}^{N}a_{ik}^2>\sum_{i=1}^{N}a_{i(k+1)}^2
\quad .
\end{equation}

We never (below) optimize with \emph{both} the smoothness
regularization and the non-negative regularization operating at the
same time.  It is left as an exercise to the reader to generalize the
non-negative g-step of \equationname~(\ref{eq:gstep_nonneg}) to the
doubly regularized case.

As a last comment we should point out that HMF (\equationname~\ref{eq:chi-squared}) in principle does always lead to components that are linearly independent, in the sense that no component can be strictly constructed by any linear sum of the other components.  The reason for this is that if a component is in the linear subspace of the others, the optimization obtains more freedom by moving it out of the subspace.

\section{Initialization}\label{sec:initialization}

In \sectionname~\ref{sec:model} we introduced the HMF model; in
\sectionname~\ref{sec:optimization} we described how to optimize from
a sensible first guess or starting point.  Here we discuss the
initialization, which is conceptually independent of both the model
definition and the optimization choice.

An initialization consists of an initial setting for the basis
spectrum parameters $g_{kj}$.  It also consists of an initial setting
for the amplitudes $a_{ik}$, but because---given basis spectra---the
finding of these amplitudes requires only weighted least squares
(given in \equationname~\ref{eq:astep} or the non-negative version of
it given in \equationname~\ref{eq:astep_nonneg}), initialization can
be thought of as only being about finding a good initial guess for the
$g_{kj}$.

There are four natural (to us, anyway) choices for initialization: $K$
spectra can be chosen at random from the data sample of $N$ spectra;
$K$ linearly independent mathematical basis functions such as
Chebyshev Polynomials or sines and cosines can be used; $K$ principal
components can be generated by a PCA; or $K$ cluster centers can be
found with a K-means algorithm run in spectrum space.  In our tests,
PCA-initialized and K-means-initialized optimizations almost always
out-performed random-spectrum and basis-function initializations.  For
this reason we only consider PCA and K-means in what follows 
(even though in the present work we make use of K-means, 
the diffusion K-means method seem to perform better in high dimensions; \citealt{richards}).

\paragraph{PCA initialization}
In what we describe as ``PCA initialization'' for $K$ basis spectra,
we in fact generalize slightly the PCA to produce a mean spectrum and
then the $[K-1]$ top eigenvectors \emph{orthogonal} to it. 
That is, before performing PCA, we \emph{project} the spectra into a 
subspace orthogonal to the mean spectrum, by scaling them 
and subtracting the mean spectrum from each one of them.
The initialization we use is the mean spectrum
and the top $[K-1]$ eigenvectors from the PCA in the orthogonal
subspace.  This methodology is rarely followed, but it is the only
method that makes sense if \textsl{(a)}~the mean is permitted to be
non-zero, \textsl{(b)}~the mean spectrum is going to be considered an
eigenspectrum or component for subsequent fitting, and
\textsl{(c)}~linear independence is important, which it always is.

In detail, we did one more ``conditioning'' step before any of this,
which was to re-scale the $M$-dimensional space (the space in which
the $\vec{f}_i$ live) so that the mean of the $N$
variances (across spectra $i$) in each coordinate $j$ was
equal.  That is, we ``isotropized'' the space from the point of view
of the observational uncertainties.  We \emph{only} performed this
isotropization for the PCA, \emph{not} for the HMF, because the HMF
takes the observational uncertainties into account naturally and
correctly.  After the mean-subtracted PCA was performed, we re-scaled
the results back to the original $M$-dimensional space for use.  This
scaling and re-scaling step is also rarely done, but must be if PCA is
to return results that are not likely to be strongly affected by
observational noise; this step effectively shrinks to small those
dimensions where the variance in the sample are likely to have been
dominated by measurement noise.

\paragraph{K-means initialization}
For the case that non-negative constraints are applied to the
estimated components and coefficients, we are forced to use an
initialization with non-negative pixel values. Even though we could
use any set of non-negative basis functions, the best initialization
seems to be the results of the K-means algorithm. That is expected
since the components extracted by K-means, that correspond to the
centers of the groups into which the algorithm divides the training
spectra, include more physical information than other non-negative
initializations. We expect that the K-means results will include only
positive values because if the number of groups is not very large
(this might result in groups with very few spectra as members), the
mean spectrum in each group very rarely includes negative values of
flux caused by errors.

\section{Applications}\label{sec:applications}
To assess the power of the technique, we are going to confirm known
double-redshift objects in the \SDSS\ spectroscopic sample. More
specifically, by using the method presented above, we define a small
number of components that is sufficient for modeling \SDSS\ 
spectra. Using these components we fit each observed spectrum at the
redshift provided by \SDSS. Then we repeat the fitting, but this time
using one set of components at the \SDSS\ redshift and one set of
components at values of redshift that lie on a nominal grid. If a
second object is present we expect the fit to be significantly improved when we use
two sets of components, one at the redshift of \SDSS\ and one at that of
the second object. In the examples that follow we will demonstrate
this using the SLACS sample of gravitational lenses \citep{bolton} and
the four known candidates of massive Black Hole Binaries (BHBs)
\citep{komossa,bogdanovic09,dotti09,boroson,shields,decarli}
\footnote{A new candidate discovered by \cite{barrows11} appeared in the literature 
only after the completion of this work. In addition, recently two systematic searches 
for black hole binary candidates in \SDSS\ spectroscopic sample were performed using HMF 
\citep{tsalmantza} and PCA components \citep{eracleous}.}.

\subsection{Training}\label{sec:training}
In order to detect the presence of two objects at different redshifts
in the \SDSS\ spectra, we need to be able to model the spectra of all
types of objects that have been observed by the survey (the primary
sample of ``Main Galaxies'', the luminous red galaxy sample or
``LRGs'', and color-selected QSOs). To do so, we have to train our
method separately for each class. For this purpose we selected a small
random sample of spectra for each type of object (approximately 5000
for Main Galaxies and LRGs and 10000 for QSOs; the numbers were
selected as such in order to make sure that we have at least as many
objects as number of pixels, which was needed for some of the tests we
performed using PCA). The spectra were taken from the 7th Data Release
(DR7) of \SDSS\ (\citealt{sdssdr7}). The values of redshift were selected to
be in the ranges of $0.01<z<0.06$, $0.20<z<0.50$ and $0.10<z<1.50$ for
Main Galaxies, LRGs and QSOs respectively.

For the selection of the LRG spectra used in this study we followed the
target selection of the LRG sample in \SDSS\ which is based on 2 cuts,
defined using magnitudes, colors and surface brightness criteria
\citep{eisenstein}. These criteria are suitable for
redshifts in the range of values greater than 0.15 and less than 0.55. For 
this study we selected galaxies that meet these criteria in the redshift range 
from 0.2 to 0.5 (95,833 sources). 

The selected samples were used to determine the maximum wavelength coverage 
for each type of object, that
is, a wavelength area for which at least 10 sources have valid data at
the bluest and the reddest wavelengths. In this way we are able to
fit the part of the spectrum that is produced by the second object for
a large range of redshift values. During this procedure, pixels with
any of the flags: SP\_MASK\_FULLREJECT, SP\_MASK\_NODATA,
SP\_MASK\_BRIGHTSKY, SP\_MASK\_NOSKY or pixels that correspond to zero
noise were treated as masked. All the spectra were
moved to the rest-frame by keeping the energy constant in each
spectral bin while relabeling the wavelength axis.  The final
wavelength coverage for each object is:
$3580.964<\lambda<9109.615\,\AA$ for Main Galaxies,
$2544.486<\lambda<7615.528\,\AA$ for LRGs and
$1522.299<\lambda<8352.183\,\AA$ for QSOs, corresponding to 4056, 4762
and 7394 pixels respectively.

For each spectrum $i$ , we obtain the official \SDSS\ pixel flux densities $f_{ij}$
interpolated by cubic-spline interpolation onto a common rest-frame
wavelength grid, logarithmically spaced in wavelength.  We also obtain
and interpolate the officially reported flux error variances
$\sigma_{ij}^2$.  Since only 10 spectra include data at the bluest
wavelengths and only (a different) 10 include data at the reddest
wavelengths, and because there are cosmic rays and other masked data
artifacts, missing data are present in all the spectra of our
sample. To deal with this problem we have linearly interpolated the 
fluxes at the masked areas, while at the missing edges we have set the fluxes 
equal to the first or the last non-masked pixel of the spectrum and 
we have set the noise of all those pixels to a very high value
($10^{-12} erg/sec/cm^2/$\AA), so that they will not be significantly
taken into account by the method; that is, we treat missing data as
simply ``badly measured'' data so as to keep the method as
straightforward as possible.

\begin{figure}
\includegraphics[angle=-90,width=0.49\columnwidth]{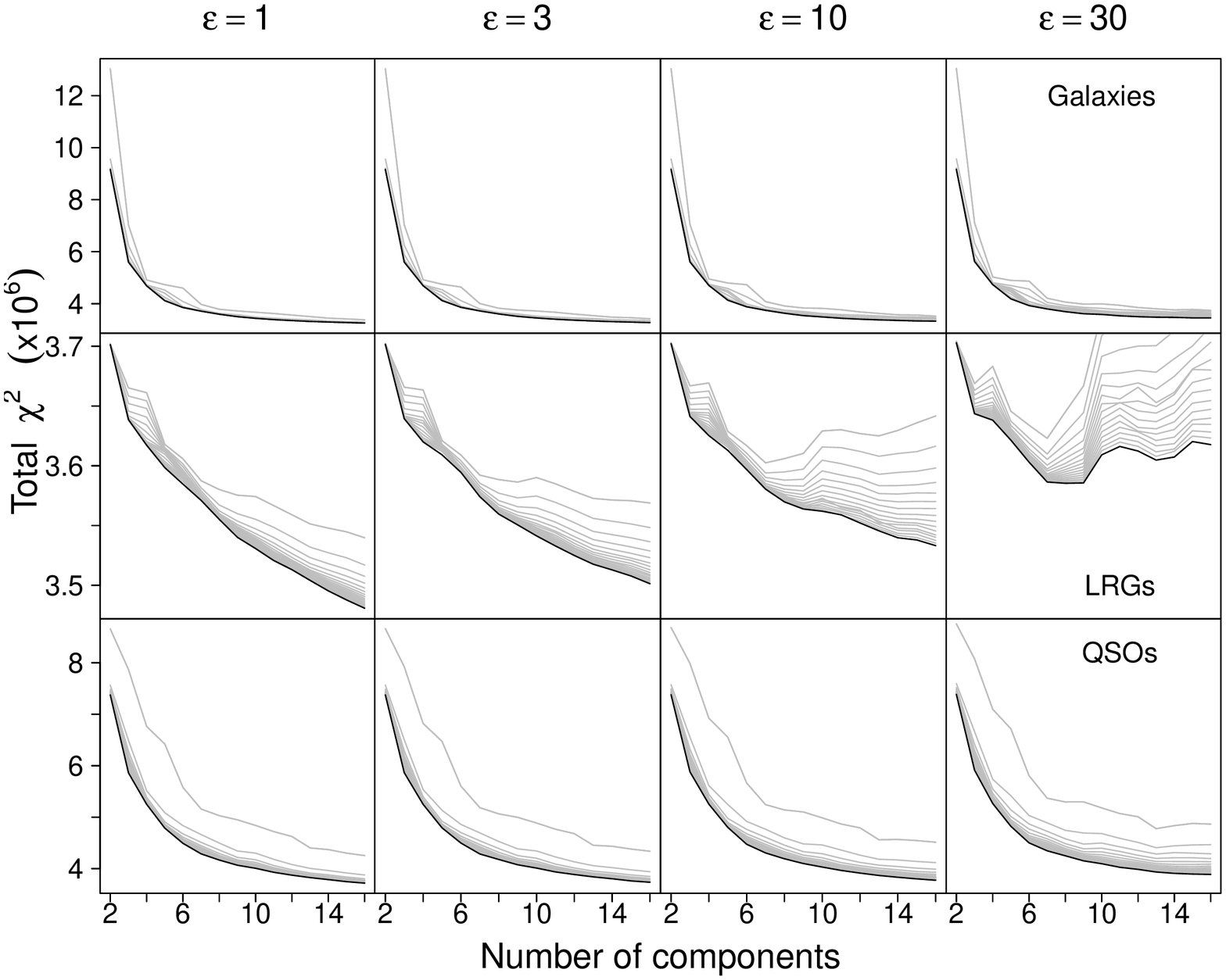}
\caption{The total $\chi^2$ value estimated by the fit of the training
  set of spectra by the components produced by the method vs. the
  number of components for each type of object (rows) and values of
  $\epsilon$ (columns).}
\label{fig:1}
\end{figure}

Using a number of spectra equal to the number of pixels selected for
each source, we performed PCA on those data. As described in 
\sectionname~\ref{sec:initialization}, the training data were
first projected into a hyperplane orthogonal to the mean
spectrum that is passing from the zero point. To do so each spectrum 
was scaled appropriately and the mean spectrum was subtracted from it. 
Additionally, the flux in each spectral bin was divided
with the RMS of the error in that pixel for all the non-masked pixels in
the training sample. The PCA results were used as an initialization to
our method. HMF was trained with a subset of approximately 1000
spectra of each type, for a different number of components and for 16
iterations, which seems to be enough for the method to reach convergence. This
can be seen in \figurename~\ref{fig:1} in which we present the results of
the fitting (total $\chi^2$, that is, the sum of $\chi^2$ values over all
wavelengths and all spectra) of the 1000 spectra of our sample for a
different number of components. This test was also performed for four
different values (1,3,10 and 30) of the smoothing factor $\epsilon$.

\begin{figure}
\subfigure{
\includegraphics[angle=-90,width=0.89\columnwidth]{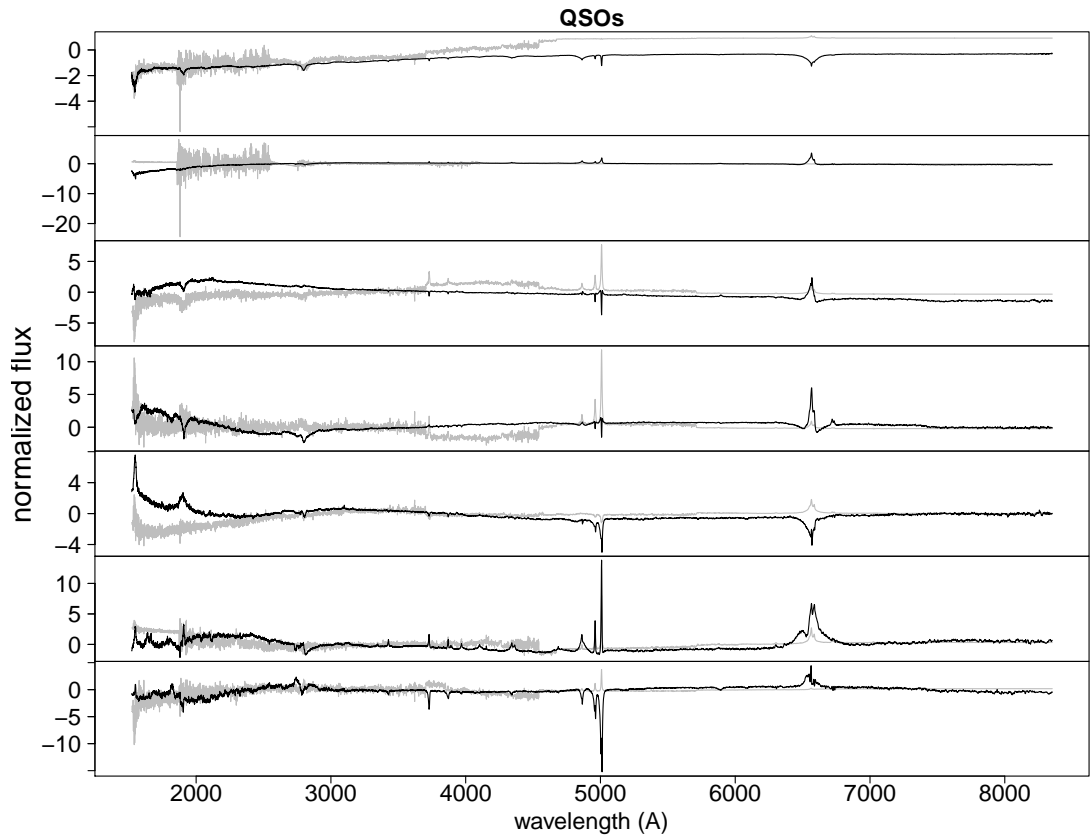}
}
\caption{The first 7 components for each type of object (Main Galaxies, LRGs and QSOs) as estimated by the PCA (grey lines) and the method presented here (black lines).}
\label{fig:2}
\end{figure}

\begin{figure}
\includegraphics[angle=-90,width=0.49\columnwidth]{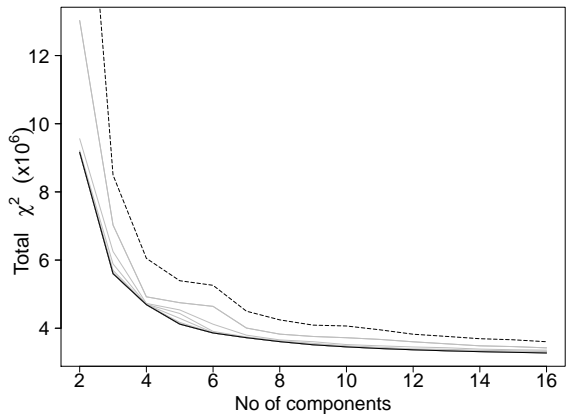}
\caption{The total $\chi^2$ value (sum over all spectra and all pixels) estimated by the fit of a test set of \SDSS\ galaxy spectra ($\approx$ 1000 objects) by the HMF (grey and black lines) and the PCA (dashed line) components vs. the number of components used in each case. From the top grey line to the bottom black one, the lines correspond to the results based on different successive iterations of the HMF (i.e. from iteration 1 to 16). It is clear that the HMF components can achieve a better fitting of the observations even from the first iteration of the method.}
\label{fig:2b}
\end{figure}

In \figurename~\ref{fig:1} we can see that the fitting of the spectra
improves a lot even after the first iteration, indicating that for a 
given number of components HMF can achieve a better modeling of the 
spectra than PCA (i.e. under the assumption that the likelihood is the best 
scalar, for a given number of components HMF can achieve a higher-likelihood than PCA). 
In \figurename~\ref{fig:2} 
we present our new set of components plotted over the
initial PCA components. We should point out that a straight comparison
between the components extracted by the two methods is not meaningful
since they span different subspaces of the observed data. A 
comparison between the methods can be achieved only by using
the results of the fitting to a set of spectra. An example of such a 
comparison is shown in \figurename~\ref{fig:2b} where both PCA and HMF 
components are used to fit the same set of \SDSS\ galaxy spectra.

In \figurename~\ref{fig:1} we also present how the components and the
total $\chi^2$ value change with the value of the smoothing factor
$\epsilon$. As was expected, by increasing the value of $\epsilon$ we 
impose more constraints, which leads to smoother components at the 
cost of less accurate fits. 

\begin{figure}
\includegraphics[angle=-90,width=0.49\columnwidth]{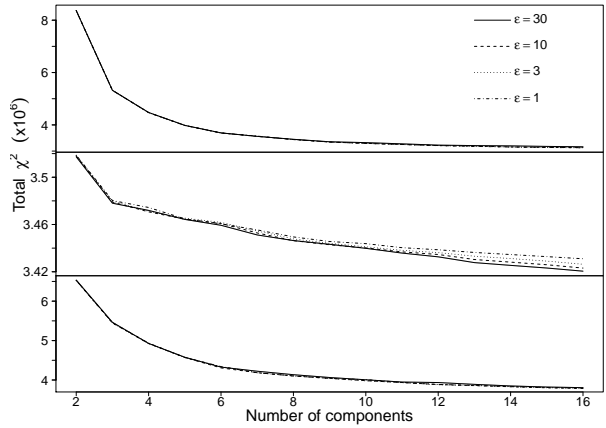}
\caption{The total $\chi^2$ value estimated by the fit of the test set
  of spectra by the components produced by HMF vs. the number
  of components for each type of object (rows: Main Galaxies, LRGs and QSOs
  respectively). The different types of lines represent the different
  values of $\epsilon$ used in each case.}
\label{fig:3}
\end{figure}

In the results presented above (\figurename~\ref{fig:1}) we have used
the same set of data to train as well as to test the method. As a
first step towards cross-validation we used a new random set of 1000
spectra as a testing set. The results of the fitting of this set, at
each iteration, with the extracted HMF components (when HMF is 
trained with the same set of spectra as before), are presented in
\figurename~\ref{fig:3}. In these plots with different types of line
we present the results for different values of $\epsilon$. As we can
see in the new test set, the best fit is achieved by different values
of $\epsilon$ and not for the smallest one as before.

\begin{figure}
\includegraphics[angle=-90,width=0.49\columnwidth]{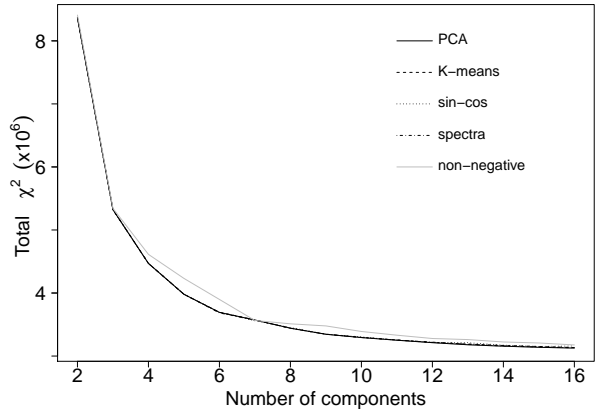}
\caption{The total $\chi^2$ value estimated by the fit of the test set
  of Main Galaxy spectra by the components produced by HMF vs. the
  number of components. The different types of lines represent the
  different initializations used (PCA, K-means, random spectra and sin
  and cosin functions).}
\label{fig:4}
\end{figure}

In order to check how much the method depends on the initialization we
repeated our tests using different sets of components as our initial
basis. In \figurename~\ref{fig:4} we present the results of the
fitting for the same test set as before when a random set of
spectra, the output of the K-means algorithm and a set of sin and
cosin functions were used to initialize the training of HMF. More specifically, in the
case that a set of random spectra was used, we selected the ones that
include data at the reddest or the bluest wavelengths. In the case of 
the K-means initialization, we used the
algorithm kmeans(stats) implemented in R with a number of centers
equal to the number of components. The algorithm uses a random
set of points as its initialization and therefore the results are
slightly different in every run.

By comparing these results we see that they are becoming worse as we
change the initialization from the PCA output, to the random spectra,
to the K-means results and to the sin and cosin functions. This result
is expected since the PCA results are chosen in a way to increase the
percentage of the total variance they include. The most probable
reason why the random spectra seem to be a better initialization than
the K-means output is that we have chosen spectra that include
information at the ends of the wavelength coverage, something that is
probably not true in the K-means initialization where the centers are
defined mainly by spectra with constant values in these areas. The sin
and cosin functions lead to the worst results as expected since they
include the least information compared to all the other
initializations used here.

\begin{figure}
\includegraphics[angle=-90,width=0.49\columnwidth]{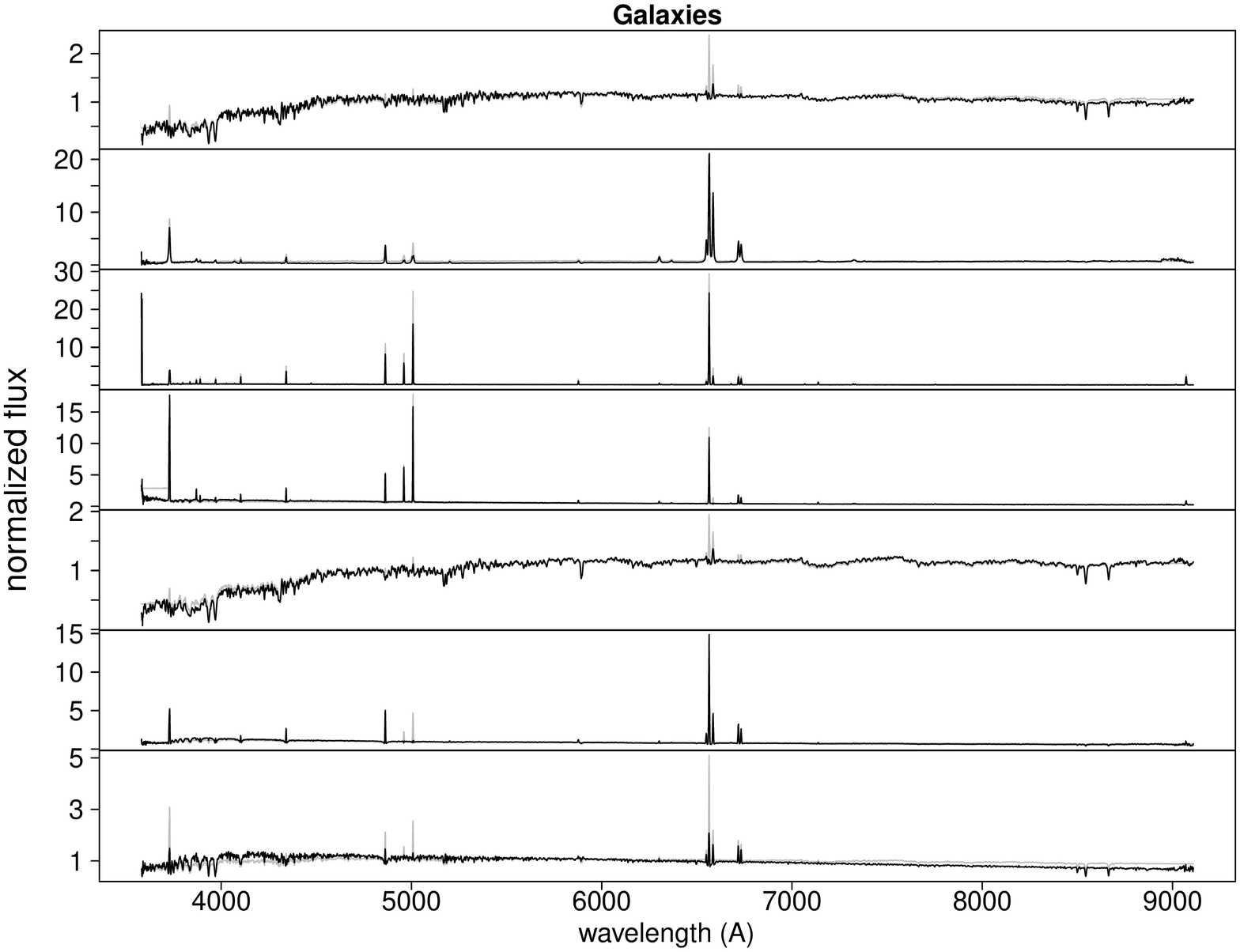}
\includegraphics[angle=-90,width=0.49\columnwidth]{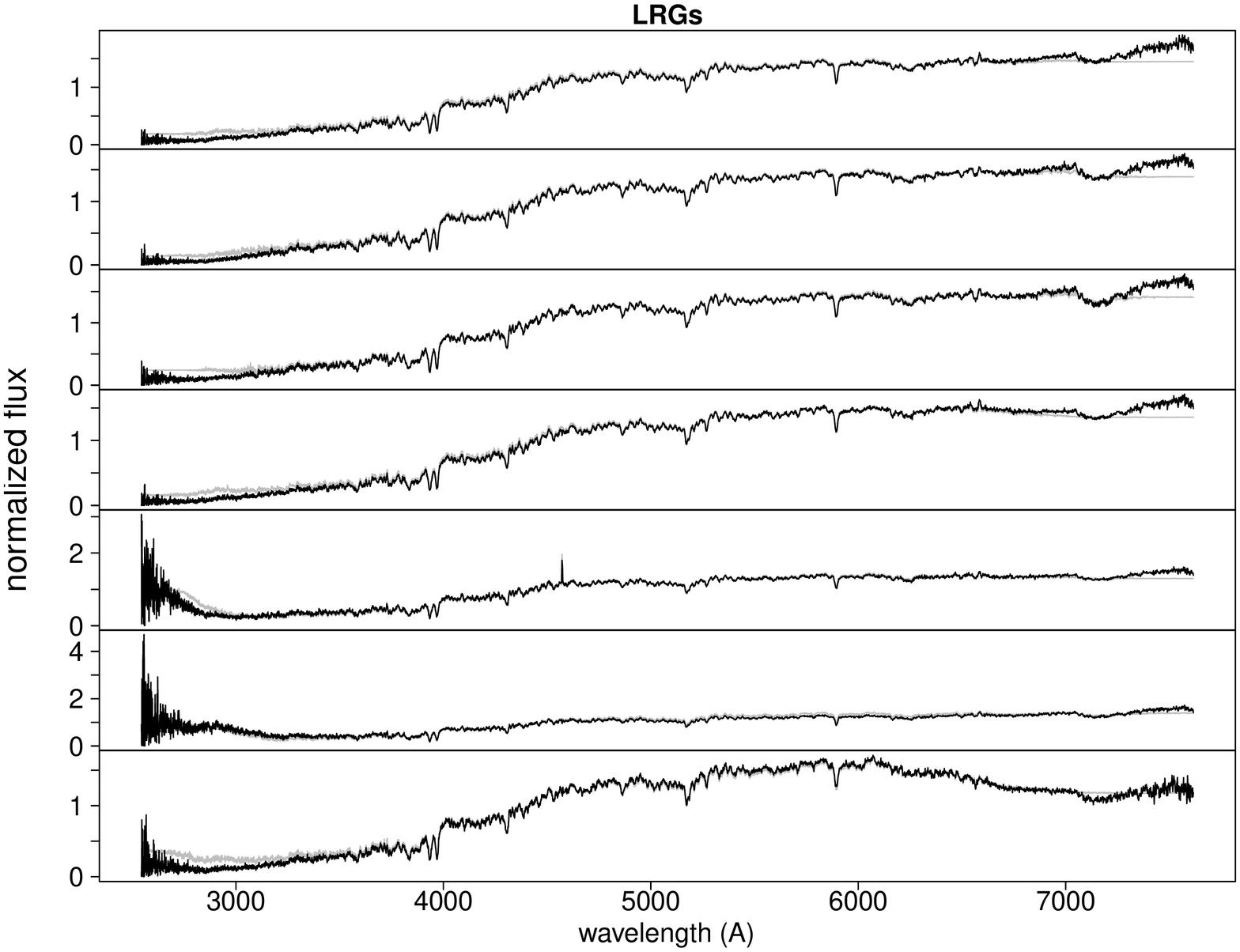}
\includegraphics[angle=-90,width=0.49\columnwidth]{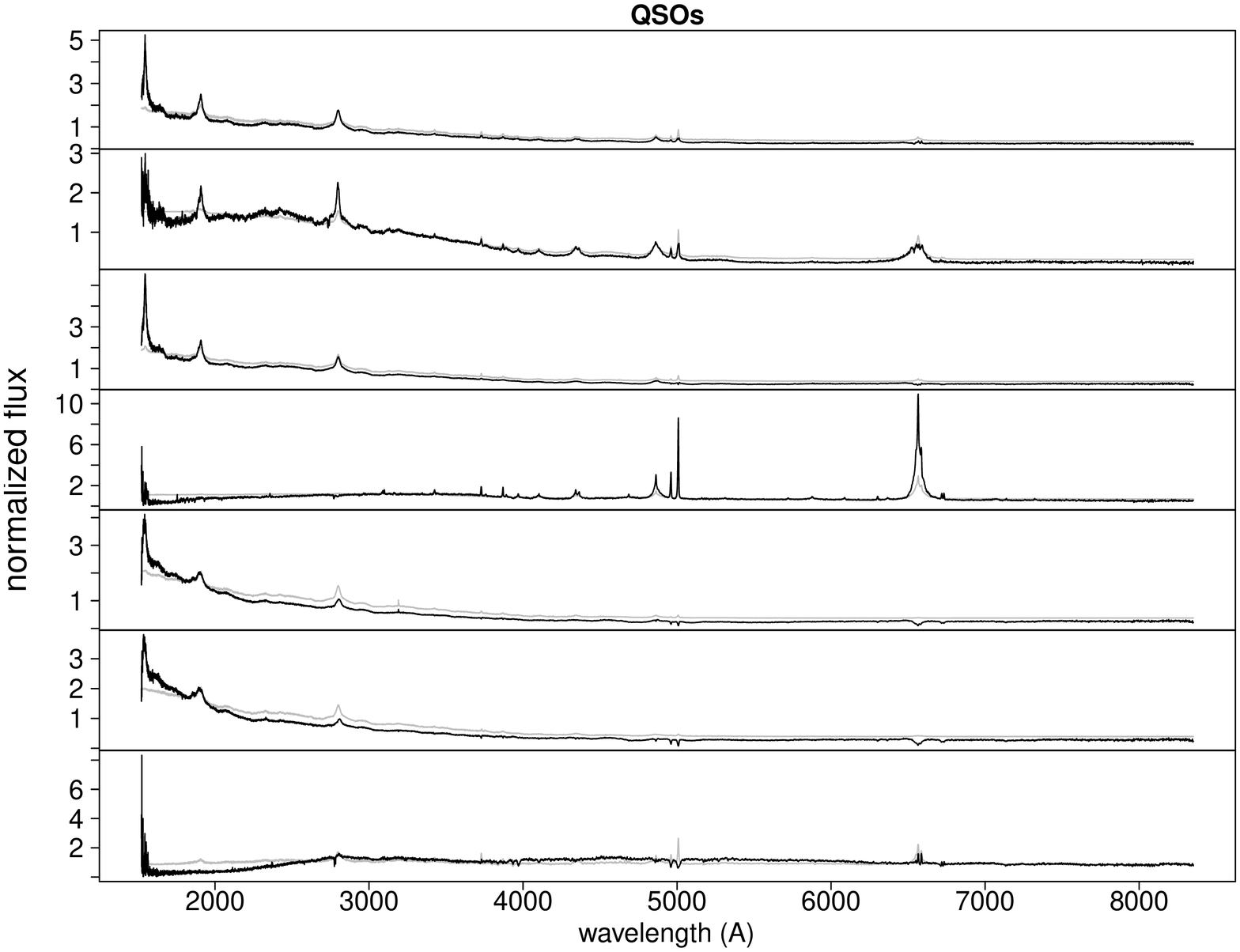}
\caption{The first 7 components for each type of object (Main Galaxies,
  LRGs and QSOs) as estimated by the K-means (grey lines) and the method
  presented here (black lines) when non-negative constraints were
  applied.}
\label{fig:5}
\end{figure}

As a last test of the method we checked how a non-negative constraint
affects the results. Since negative values in the spectra are caused
by observational errors, modeling the spectra of astronomical sources
with components that include negative values has no physical
meaning. This problem can be solved be applying a non-negative
constraint to our basis. The way that this is achieved is by
initializing with a non-negative set of components and coefficients
and iterating according to the non-negative a-step of
\equationname~(\ref{eq:astep_nonneg}) and the non-negative g-step of
\equationname~(\ref{eq:gstep_nonneg}). One of the best ways to
initialize this method with a set of non-negative components that
include physical information is to use once again the K-means
algorithm. The results of the fitting of the test set of spectra with
the components extracted in this way after 2048 iterations are
presented in \figurename~\ref{fig:4}, while in \figurename~\ref{fig:5}
we present the resulting components for each type of object for $K=7$.

By comparing those results with the ones obtained without the
non-negative constraint we see that the fitting is now worse. This was
expected since non-negativity is a very strict constraint. On the
other hand even if the components now seem to have a better physical
meaning, that is, they look more like spectra of particular type of
objects, in many cases there seems to be a problem at the edges of the
spectra where they tend to start from exactly zero values (for
example, the components for LRG spectra). At this point we should
mention that when applying the method for the non-negative case we
have not used an additional smoothing constraint.

Based on the results presented here and an additional test, that is 
defining the minimum number of components required to detect the 
second redshifts in the SLACS and BHB candidate samples, we selected 
the optimal number of components and value of $\epsilon$. More specifically, 
we decided to fit the \SDSS\ spectra using the 14, 7 and 14 components 
that were produced by HMF when trained with Main Galaxy, LRG
and QSO spectra for 16 iterations and for $\epsilon$=3, 30 and 10
respectively. It is interesting that fewer components are needed in
order to fit well the LRG \SDSS\ spectra than the Main
Galaxy and QSO spectra.  This is expected, since the LRG spectra show
very little variation \citep{eishogg} in comparison to other 
types of sources like QSOs \citep{yip}. A more detailed description of
the fitting and its results is presented below.

\subsection{Two-redshift models}\label{sec:tworedshifts}
In order to detect the presence of a second redshift in the \SDSS\ 
spectra we compute the improvement of the fitting when a spectrum is
fitted with two sets of componets at two different redshifts (that is, the
one estimated by \SDSS\ and another redshift) instead of only one set of
components at the redshift provided by \SDSS. This second redshift is
scanning a regular grid of values selected to be uniform in a
logarithmic scale, that is, in the same way as the wavelengths in \SDSS. In
this way moving to the next value of the redshift grid is equivalent
to shifting the spectrum by one pixel. In practice the improvement of
the fitting can be estimated by measuring the $\chi^2$ difference
between the two fits ($\Delta\chi_{in}^2$):
\begin{eqnarray}\label{eq:deltachi-squared}\displaystyle
\Delta\,\chi_{in}^2 & = & \sum_{j=m}^{M'}
             \frac{\left[f_{ij}-\sum_{k=1}^K a_{ik}
                      \,g_k(\lambda_j/[1+z_i])\right]^2}
{\sigma^2_{ij}} - \nonumber\\ 
& - & \sum_{j=m}^{M'}
             \frac{\left[f_{ij}-\sum_{k=1}^K a'_{ik}
                      \,g_k(\lambda_j/[1+z_i])-\sum_{k=1}^K \beta_{ik}
                      \,g_k(\lambda_j/[1+z_n])\right]^2}
{\sigma^2_{ij}}
\quad ,
\end{eqnarray}
where $m$ and $M'$ are the first and last common pixels between the
components when moved to the \SDSS\ and the second redshift ($z_i$ and
$z_n$ respectively). By definition an improvement of the fitting will 
result in a positive value of $\Delta\chi_{in}^2$. In the case that 
the spectrum has significant flux coming from a second object, 
we expect that there will be a peak in $\Delta\chi_{in}^2$ at a second 
redshift equal to the one of that object. The strength of the peak depends 
on the brightness and therefore the distance of the second object, the 
presence of emission lines in its spectrum and to some extent to the 
spectral coverage. As it is obvious the use of an additional set of 
componets always improves the fitting of the spectrum. However, this 
improvement is not significant and does not vary a lot for different 
values of the second redshift in the cases that the signature of a second 
object is not present in the spectrum. Finally, since we perform our search 
using the difference of the $\chi^2$ values between the two fits, values of
$\Delta\chi_{in}^2$ that correspond to different second redshifts are
directly comparable, despite the fact that they might correspond to
different number of pixels in the fits.

\subsection{Testing}

\subsubsection{The SLACS survey}
The SLACS survey \citep{bolton} includes 131 strong galaxy-galaxy
gravitational lens candidates, selected by the presence of higher
redshift emission lines on the top of a lower redshift stellar continuum. 
Using the components extracted by the method presented here
(\sectionname~\ref{sec:training}) for Main Galaxy and LRG \SDSS\ spectra
we applied the test described above
(\sectionname~\ref{sec:tworedshifts}) in order to reproduce the results of the
SLACS survey for the second redshift.

As a first step we applied the test using the 14 Main Galaxy components of
\sectionname~\ref{sec:training} for both the foreground and the
background object. For each spectrum of the SLACS survey we used 
the procedure described in \sectionname~\ref{sec:tworedshifts} to
search for peaks of $\Delta\,\chi_{in}^2$ corresponding to the second redshift. 
An additional criterion that we used was that the peaks (if present) should 
correspond to fits that did not produce negative [OIII] lines if they were 
included in the spectral range of the fit. In this way we detected peaks for 119 SLACS 
spectra at the same redshifts as those found in the SLACS survey 
(for an example, see \figurename~\ref{fig:6}).

For the remaining 12 cases we applied the same procedure but this
time using the LRG components to fit the foreground object and the
Main Galaxy components to fit the background one. The results show that
using this approach we were able to extract the same results as the SLACS
survey for 6 of those spectra (an example is shown in \figurename~\ref{fig:7}).

In the other 6 cases we detect a different second redshift than
the one reported by \cite{bolton}. For those objects we applied once
again our method but this time assuming the presence of three objects
instead of two. More specifically, we fit the spectrum using a set of
components (LRG or Main Galaxy) at the \SDSS\ redshift, a set of components
(Main Galaxy) at the second redshift to which was given the highest
probability by our method, and a set of components (Main Galaxy)
at a redshift scanning a regular grid of values. This time we managed
to predict the SLACS second redshift for 4 additional objects (\SDSS\ J1155+6237, 
\SDSS\ J1618+4353, \SDSS\ J1621+3931 and \SDSS\ J1718+6424),
(an example is shown in \figurename~\ref{fig:8}). For at least two of these 
sources (\SDSS\ J1618+4353 and \SDSS\ J1718+6424) the presence of three objects 
was confirmed by high resolution \project{HST} imaging observations which showed that in these 
two cases the lens consists of two foreground galaxies (\citealt{bolton06}; \citealt{bolton}).

\begin{figure}
\includegraphics[angle=-90,width=0.49\columnwidth]{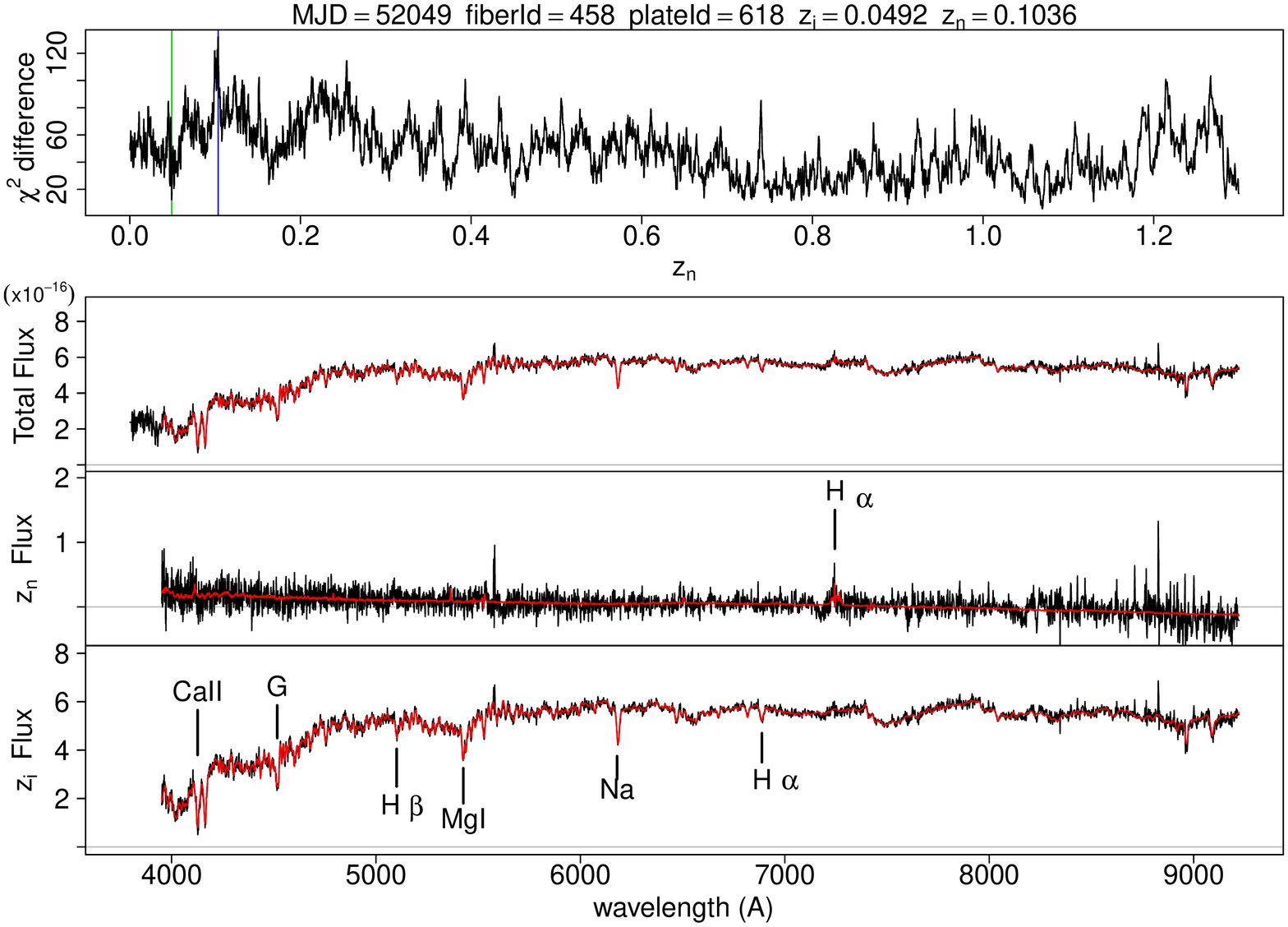}
\includegraphics[angle=-90,width=0.49\columnwidth]{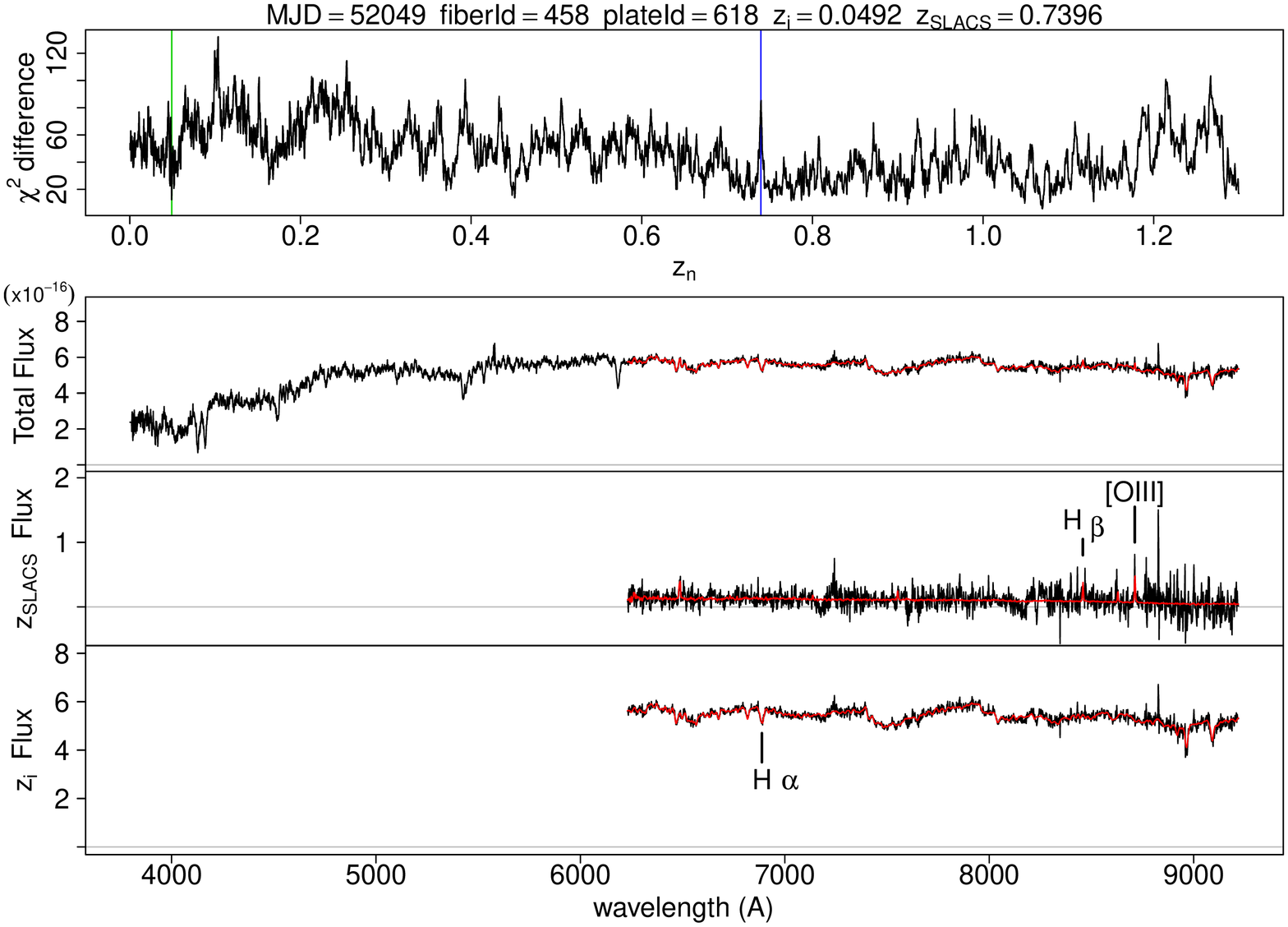}
\includegraphics[angle=-90,width=0.49\columnwidth]{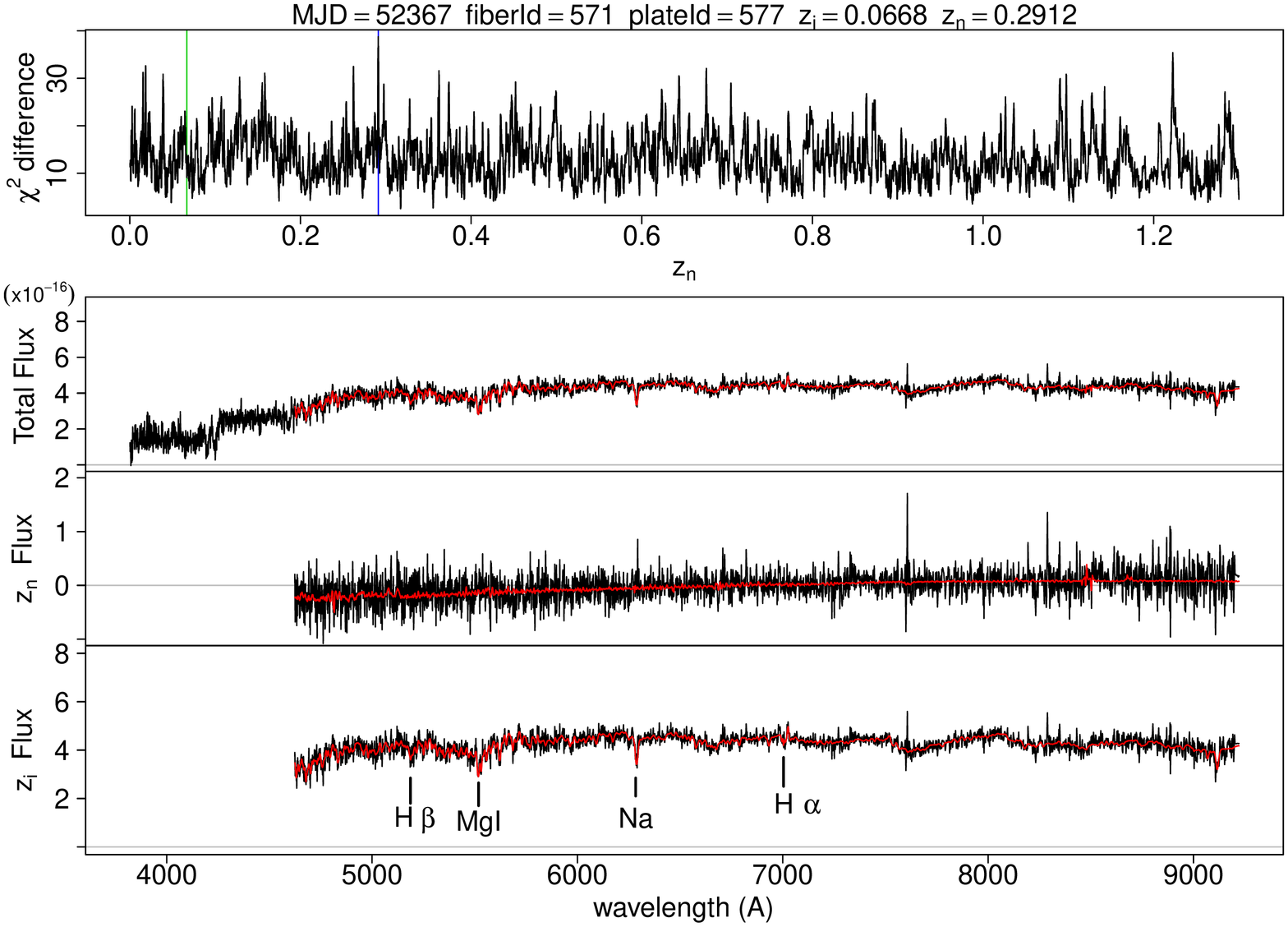}
\includegraphics[angle=-90,width=0.49\columnwidth]{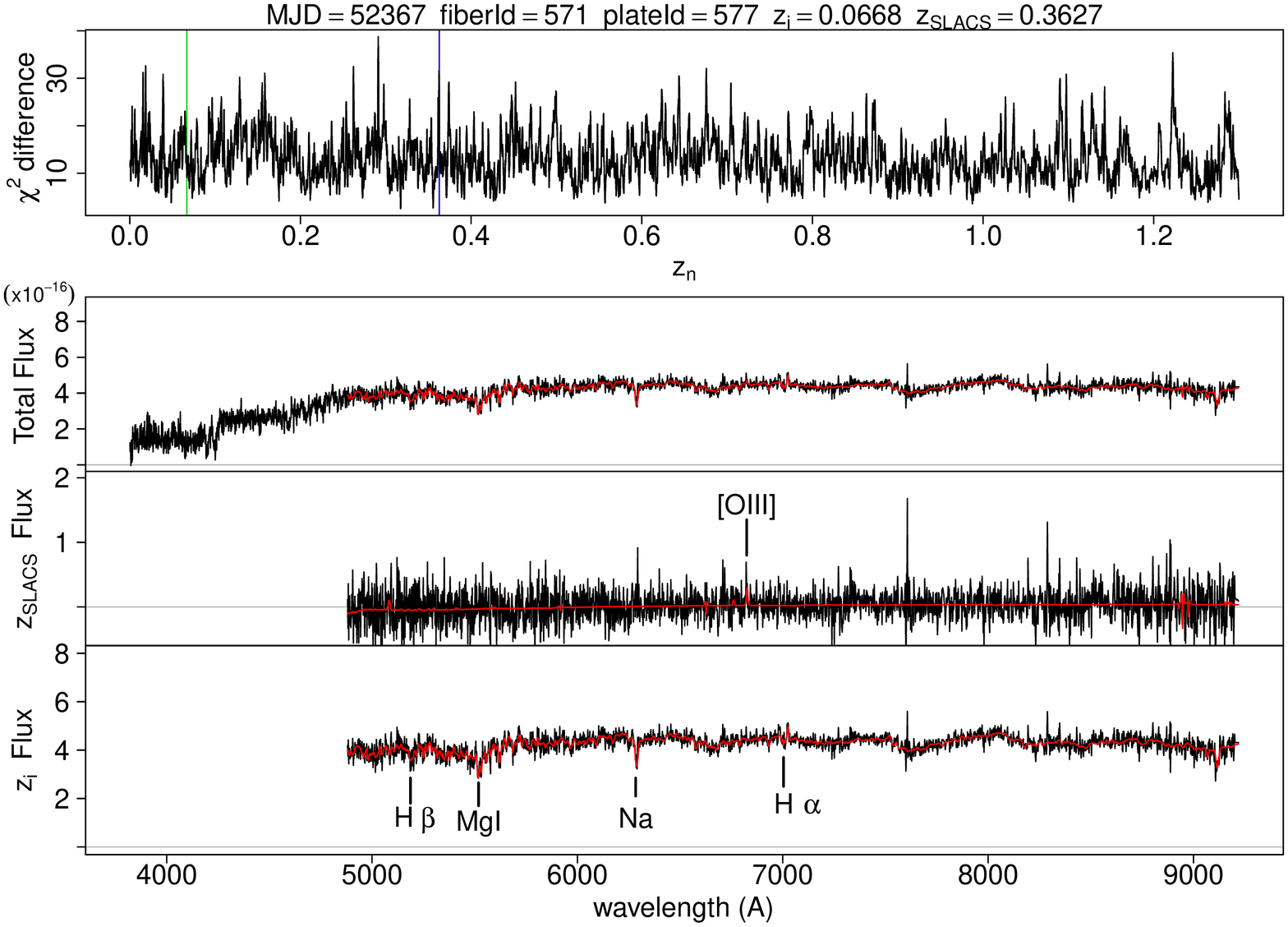}
\caption{The 2 lens candidates in the SLACS survey that could not be confirmed by the
  method presented here. The left plots show the results of the
  fitting with the second set of components at the second redshift
  estimated with our method, while the right at the SLACS second
  redshift.}
\label{fig:9}
\end{figure}

For only 2  (\SDSS\ J1039+0513 and \SDSS\ J1550+5258) out of the 131 spectra 
tested here we were not able to detect the second redshift. The results for these spectra as well as
the fitting at the second redshift given by SLACS are shown in
\figurename~\ref{fig:9}. From this \figurename\ it is clear that these are
weak detections with no obvious signature of an additional object at
the second redshift.

The results so far are very promising. Our goal is to apply this method to the
whole \SDSS\ spectroscopic sample in order to detect new gravitational
lens candidates.

\subsubsection{The known sample of BHB candidates}
Another type of object that can be detected by the presence of two
redshifts in its spectrum is the BHBs. In the case of BHBs we
expect the presence of two sets of emission lines (one broad and one
narrow) with a velocity shift between them, caused by the rotation
of the less massive black hole around the more massive one that is
located at the center of the system. Only four objects in \SDSS\ had
been selected to be BHB candidates until the completion of the 
tests presented here
\citep{komossa,bogdanovic09,dotti09,boroson,shields,decarli}. By applying our
method to those spectra we followed the same procedure as in the case
of gravitational lenses in order to test if we can detect the second
redshift. However, since in this case the separation between the two
sets of lines is expected to be small, we limited our search to
redshift differences below 0.1. The second redshift can be either 
smaller or larger than the \SDSS\ one.

\begin{figure}
\includegraphics[angle=-90,width=0.49\columnwidth]{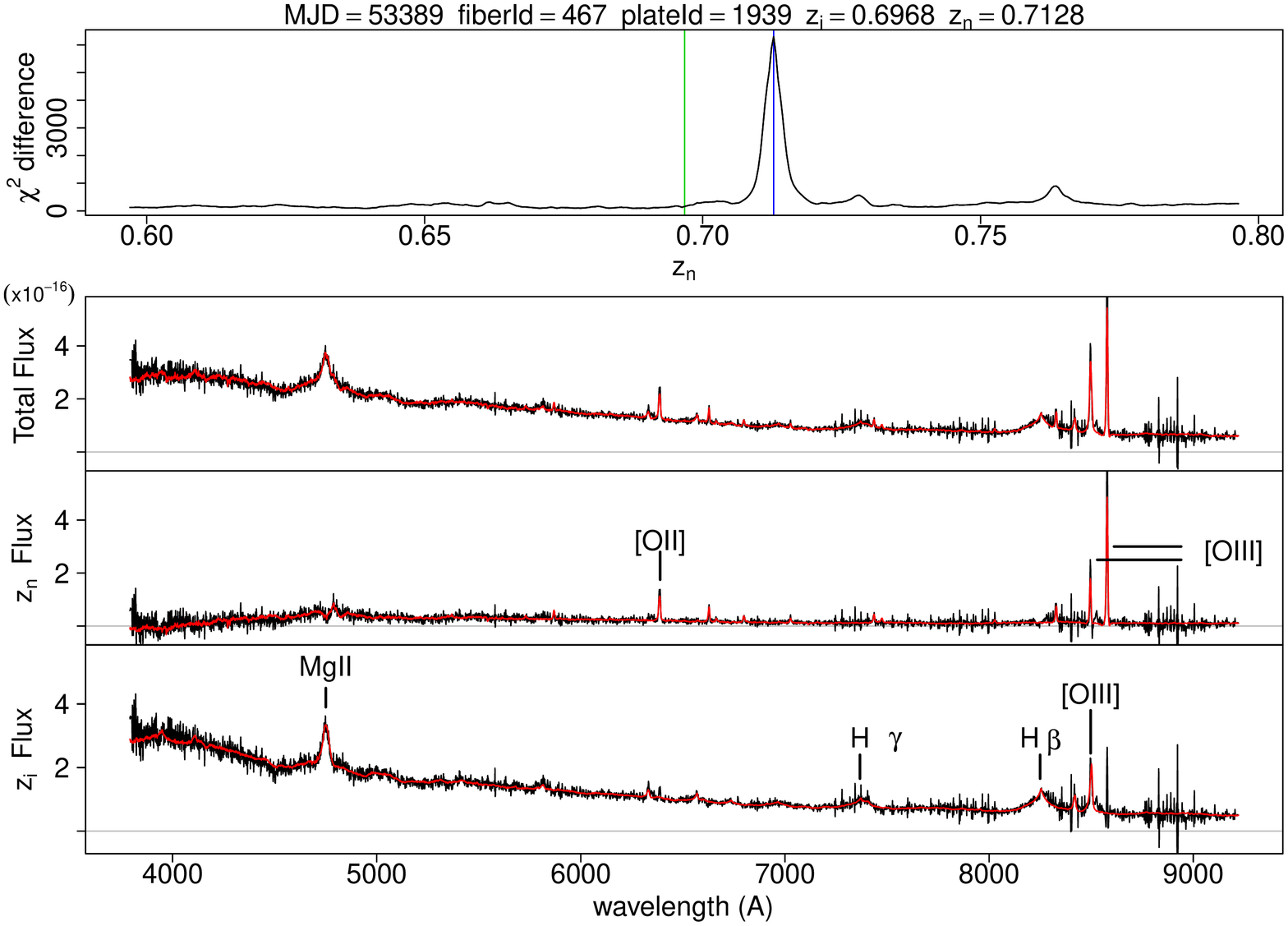}
\includegraphics[angle=-90,width=0.49\columnwidth]{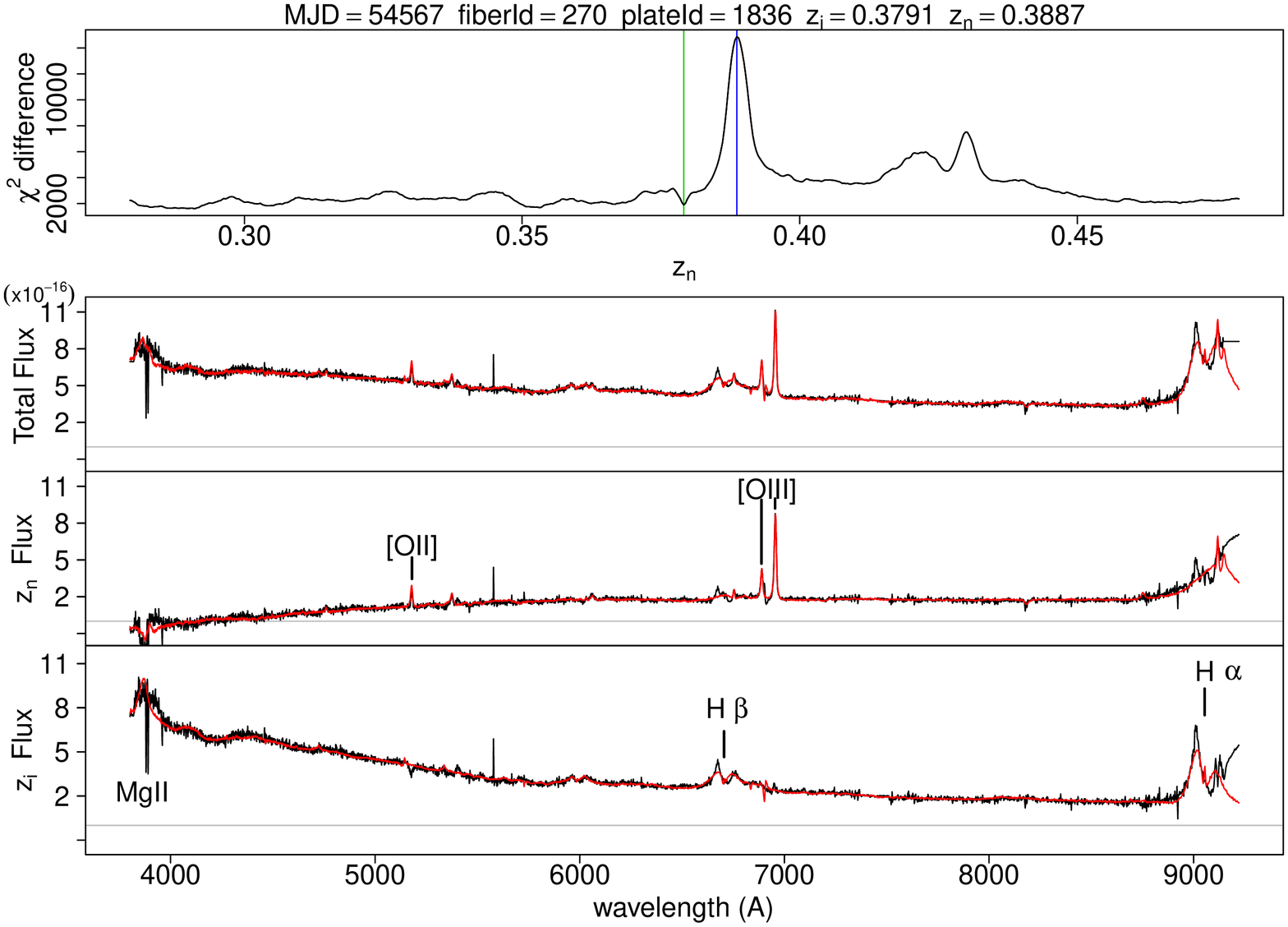}
\includegraphics[angle=-90,width=0.49\columnwidth]{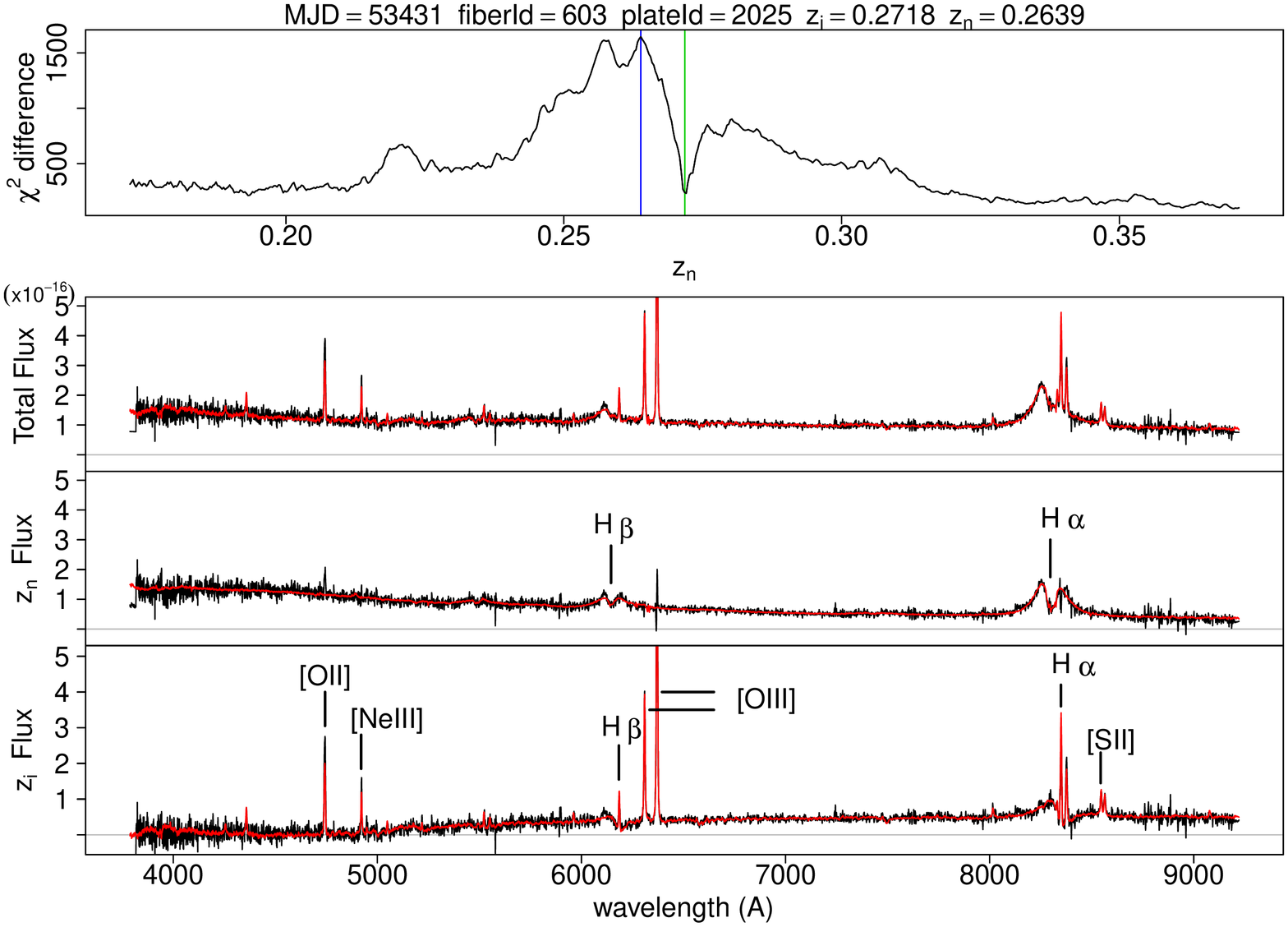}
\includegraphics[angle=-90,width=0.49\columnwidth]{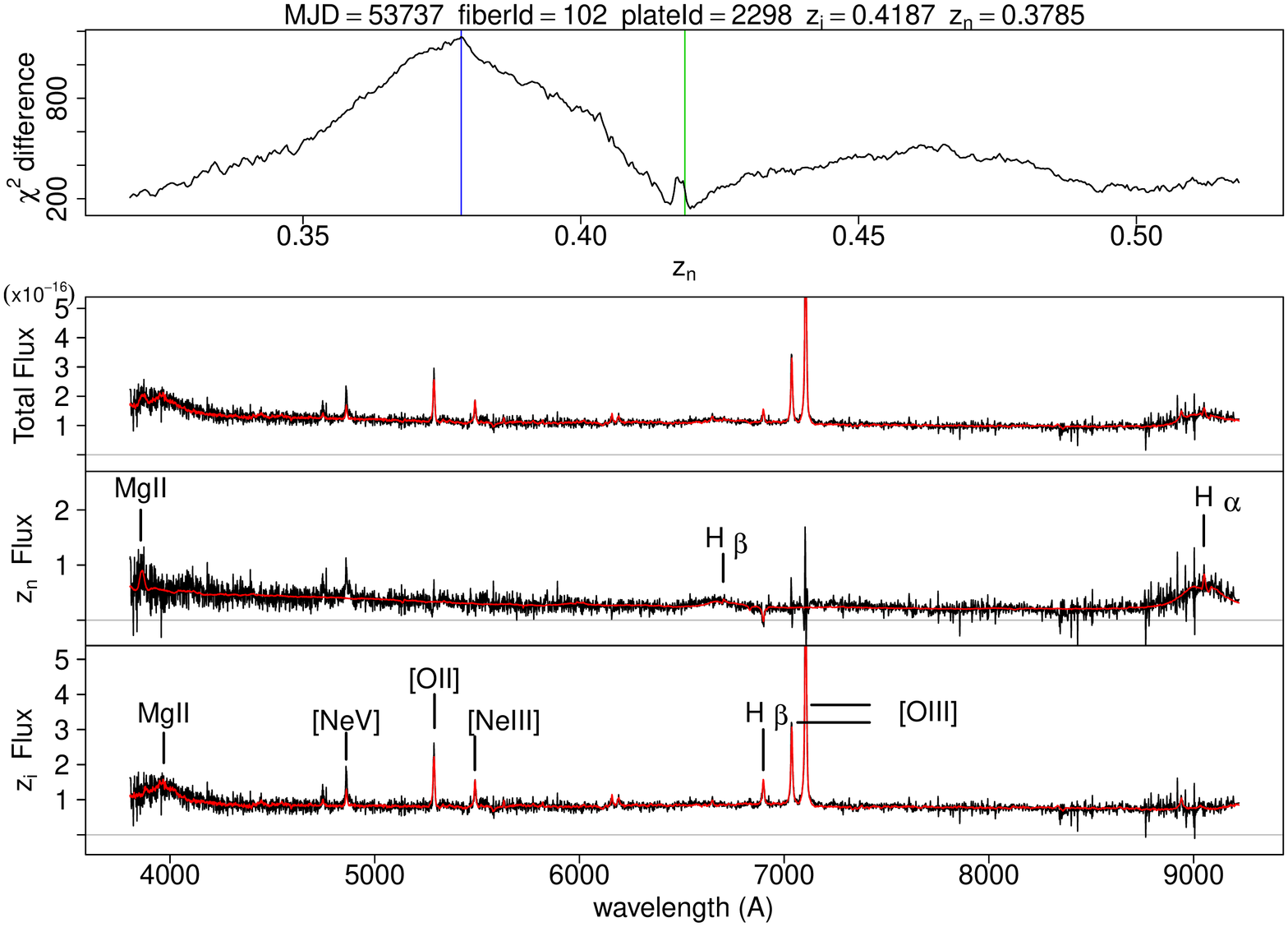}
\caption{The four known BHB candidates. \textsl{Top panel:} The
  $\chi^2$ difference between the fitting of the spectrum with 1 and 2
  sets of components. The green and blue lines represent the \SDSS\  
  redshift ($z_i$) and the one with the largest $\chi^2$
  difference ($z_n$). \textsl{2nd panel:} The fitting of the spectrum (black)
  with both sets of components (red). \textsl{3rd panel:} The part of
  the spectrum (black) fitted by the one set of components at the
  redshift with the highest $\chi^2$ difference (red). \textsl{Bottom
    panel:} The part of the spectrum (black) fitted by the one set of
  components at the \SDSS\ redshift (red). The wavelength coverage 
of the fit in each panel (red line) is defined by the common area between the SDSS spectrum and 
the 2 sets of components when moved to the $z_i$ and $z_n$ redshifts.}
\label{fig:6a}
\end{figure}

\begin{figure}
\includegraphics[angle=-90,width=0.49\columnwidth]{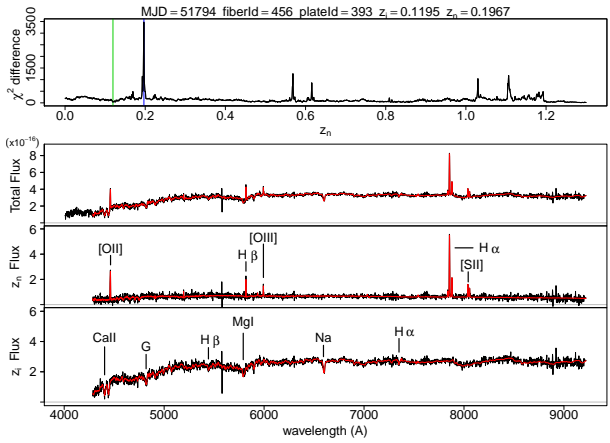}
\caption{As \figurename~\ref{fig:6a} for an example of a lens candidate in the SLACS survey
  which was fitted with two sets of Main Galaxy components.}
\label{fig:6}
\end{figure}

\begin{figure}
\includegraphics[angle=-90,width=0.49\columnwidth]{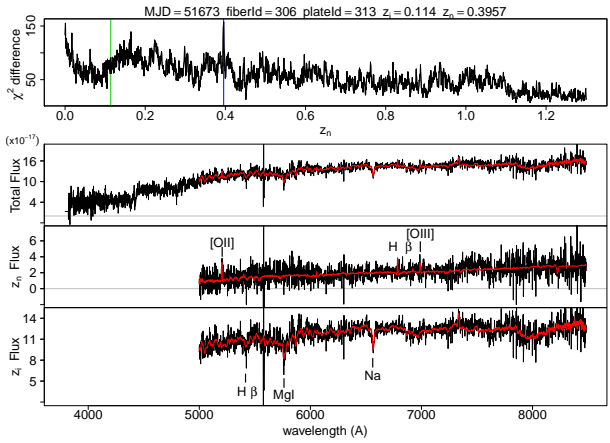}
\caption{As \figurename~\ref{fig:6a} for an example of a lens candidate in the SLACS survey
  which was fitted with one set of LRG components and one set of
  Main Galaxy components.}
\label{fig:7}
\end{figure}

\begin{figure}
\includegraphics[angle=-90,width=0.49\columnwidth]{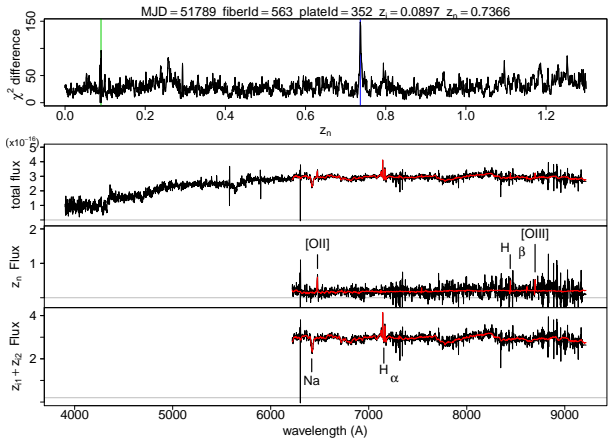}
\caption{As \figurename~\ref{fig:6a} for an example of lens candidate in the SLACS survey
  that was fitted with three sets of components (Main Galaxies).
  The bottom pannel shows the fitting of the residuals by two
  sets of components.}
\label{fig:8}
\end{figure}

The spectra were fitted using a set of QSO components that were
extracted in \sectionname~\ref{sec:training} at the \SDSS\ 
redshift and another set of the same components at a redshift scanning a
narrow grid of regular values. The results for the four candidates are
presented in \figurename~\ref{fig:6a} where we can see that we are
able to reproduce all of the four spectra with shifted broad lines given 
in the literature.

This method was applied to spectra of 54\,586 and 3929 objects
spectroscopically classified as QSOs and galaxies respectively in \SDSS\ 
DR7, with 0.1$<$z$<$1.5 in order to detect more candidates of this
type of object (\citealt{tsalmantza}). The search resulted to 32
objects with peculiar spectra, nine of which can be interpreted as BHB
candidates.

\section{Discussion}\label{sec:discussion}
We have developed a new technique called ``HMF'' for dimensionality
reduction, similar to PCA and other kinds of factor analysis, but
based on optimization of a probabilistically justifiable objective
function.  The method produces---in principle---the $K$ components
whose linear combination best reproduces well the whole training set
of the observations, given the reported observational uncertainty
variances.  Because the method makes proper use of the observational
uncertainties, it also deals properly with missing data; it does not
require that each data point has a measurement on every axis, nor does
it involve heuristic interpolation or patching of those missing data.

Since HMF is based on the minimization of a total $\chi^2$---unlike
PCA, which is based on maximization of the observed data variance
captured by the top components---it produces basis functions that fit
the real data much better than the PCA results, for the same number
$K$ of components, essentially \emph{by construction}.  In contrast to PCA,
HMF is also able to extract more information from the training set
because it uses many data points that could not be used with PCA, and
more data dimensions or directions per data point. An example of this
is that we managed to achieve maximum wavelength coverage in the
HMF-generated components even though the training set included objects
in a large range of redshifts, not one of which has data over the full
wavelength range.  PCA has one advantage over HMF, and that is
speed. Even though the individual HMF $\chi^2$ minimization iterations
steps are fast, the number of iterations make the method still much
slower than PCA. This is the price of probabilistic righteousness.

At this point it is responsible to note that the ``a-step, g-step''
formalisms presented here represent only one choice among many for
optimization: Specialists in optimization might point out that this
iterative scheme is itself heuristic, and there might be far faster
methods that could be found by good non-linear least-squares
optimization systems.  Along these lines, all derivatives of the
bilinear model are easy to compute analytically.

Perhaps the biggest advantage of HMF over PCA is that, because it
makes proper use of the errors, it does not require a high
signal-to-noise data set for training.  Because PCA models
\emph{observed variance}, at low signal-to-noise it captures the noise
rather than the signal.  HMF will not have this property, at least in
the limit of large numbers. Having said that, we should mention 
that this does not mean that intersting results cannot be obtained by 
applying PCA to low signal-to-noise data. As an example we refer to the 
\citet{borosona} study, where the authors have successfully identified a 
number of outlier sources (including a BHB candidate) using SDSS low 
signal-to-noise spectra.

We applied the HMF method to spectra from the \SDSS, building a data-driven
model of the spectra over a wide wavelength range.  This model does an
excellent job of explaining the spectra even with a small number $K$
of components.  The number of components K that is required to 
sufficiently model the data in each case, depends on the training set 
used by the method, e.g. its dimensionality $M$, the variance that it 
includes, the rank of its subspace etc. In order to choose the optimal 
number $K$ of components a kind of cross-validation was employed: The 
method is trained on a training set of spectra with different values 
of $K$; the trained components are used to fit a test set of observed 
spectra not in the training set.

For the training part of the method we use a subset of the observed
spectra to which we want to apply the results. One way to improve the
results presented here would be to use the whole set of spectra,
except the one under testing, in order to train the method. However,
this requires some engineering in the case of large surveys like
\SDSS, and for that reason we choose to use a subsample of the data
for training purposes.

Another way to improve the results would be to add prior information
on the amplitudes. Based on the coefficients extracted by the fitting
of the training spectra by the components, we can draw conclusions
about the real amplitudes that are likely to occur in the world of
real astronomical spectra.  Hierarchically inferred priors in
amplitude space would improve performance at low signal-to-noise, and
make the method more sensitive to outliers and unrealistic solutions.
However, this also requires significant engineering that goes beyond
the scope of the present work.

Another disadvantage of HMF relative to PCA is the existence of many
local optima---HMF is not convex.  This issue can be ameliorated by
using different initializations and finding the local optima that
produce the best fit to the test data.  Also, because there are
multiple optima that differ only by permutations and linear
combinations of the same basis spectra (there are subspace-description
degeneracies), comparisons among resulting components extracted from
different initializations (or different methods) shouldn't be performed
by straight comparison between the components themselves; different
solutions ought to be compared in the data space, or in the quality of
the fit to a good test set of real data.

An additional advantage of HMF over PCA is that it
also provides an option for non-negative and smoothness constraints in
the resulting components and coefficients.  This option can help
produce results that do not include unphysical features (for example,
negative emission lines or features smaller in scale than the
spectrograph resolution). We should keep in mind though that if
applied inappropriately, these constraints can have a big impact in
the results and produce a poorer fit to the observed spectra.

The model is excellent for anomaly detection:  We applied the HMF
model produced with the \SDSS\ training set to the problem of
confirming double-redshift objects.  Of the 131 galaxy--galaxy
gravitational lenses in the SLACS survey we were able to automatically
detect 129, using components trained on the \SDSS\ Main Galaxy and LRG
spectra. The confirmation was made by fitting the spectra with a
mixture of two sets of spectral model components, one at the
\SDSS\ redshift and one at a second redshift; the quality of fit was
compared to a single-redshift fits.  In a similar manner, we were able
to recover a set of four previously known black-hole binary
candidates.  In the future, we plan to perform comprehensive automatic
searches for objects of these types in the entire \SDSS\ spectroscopic
data set.

Another application of the HMF method could be for the determination
of more accurate redshifts for single objects.  This application could
be very interesting for the case of QSOs at high redshifts, where
narrow lines don't appear in the observed optical spectral domain, and for
defining template spectra for redshift estimation of objects that are
going to be observed at low signal-to-noise in future surveys; by
construction the method produces the best possible model of the
objects under study (when the training set is appropriate).

Finally, it is worth noting that nothing in the method is specific to
spectral data---it could be applied to any data for which a linear
model makes sense---and nothing is specific to the delta-function
basis of ``spectral pixels''.  The method produces a linear model;
this can be passed through any linear basis functions (as it is
in \citealt{blanton}).  Some such
transformations could lead to faster or better regularized results at
essentially no cost.

\acknowledgments It is with deep sadness that we thank our deceased
colleague Sam Roweis (Toronto, NYU), who initiated this work by
suggesting the model and all the optimization strategies.  We also
thank 
 Coryn Bailer-Jones (MPIA),
 Mike Blanton (NYU),
 Jo Bovy (NYU),
 Adam Bolton (Utah),
 Todd Boroson (NOAO),
 Roberto Decarli (MPIA),
 Massimo Dotti (MPA),
 Rob Fergus (NYU),
 Dustin Lang (Princeton),
 Zoubin Ghahramani (Cambridge University),
 Tod Lauer (NOAO),
 Nell Lawrence (University of Sheffield),
 Dmitry Malyshev (Stanford),
 Iain Murray (Edinburgh),
 Joseph Richards (Berkeley),
 Hans-Walter Rix (MPIA),
 Kester Smith (MPIA),
 Ben Weaver (NYU), and
 Andrew Wilson (Trinity College, Cambridge University) for valuable
comments and discussions. Partial funding for this project was
provided by NASA (grant NNX08AJ48G), the NSF (grants AST-0908357 and IIS-1124794), and
the Alexander von Humboldt Foundation.  This research made use of the
SAO/NASA \project{Astrophysics Data System} and the \project{R}
Project for Statistical Computing.  All the code used in this project
is available from the authors upon request.

Funding for the Sloan Digital Sky Survey (SDSS) has been provided by
the Alfred P. Sloan Foundation, the Participating Institutions, the
National Aeronautics and Space Administration, the National Science
Foundation, the U.S. Department of Energy, the Japanese
Monbukagakusho, and the Max Planck Society. The \SDSS\ Web site is
http://www.sdss.org/. The \SDSS\ is managed by the Astrophysical
Research Consortium (ARC) for the Participating Institutions. The
Participating Institutions are The University of Chicago, Fermilab,
the Institute for Advanced Study, the Japan Participation Group, The
Johns Hopkins University, the Korean Scientist Group, Los Alamos
National Laboratory, the Max-Planck-Institute for Astronomy (MPIA),
the Max-Planck-Institute for Astrophysics (MPA), New Mexico State
University, University of Pittsburgh, University of Portsmouth,
Princeton University, the United States Naval Observatory, and the
University of Washington.

{}

\clearpage

\clearpage

\begin{figure}
\subfigure{
\includegraphics[angle=-90,width=0.89\columnwidth]{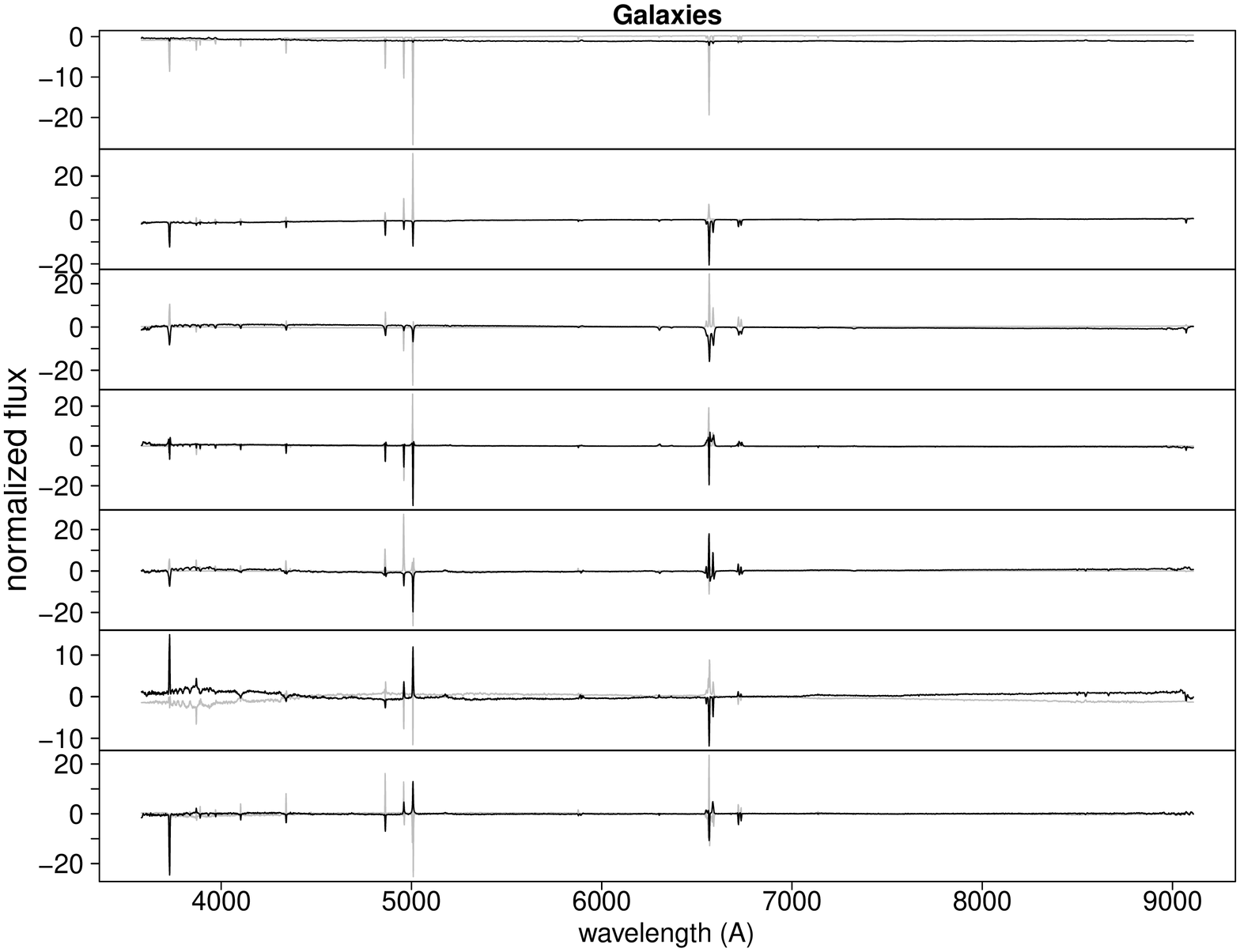}
}
\subfigure{
\includegraphics[angle=-90,width=0.89\columnwidth]{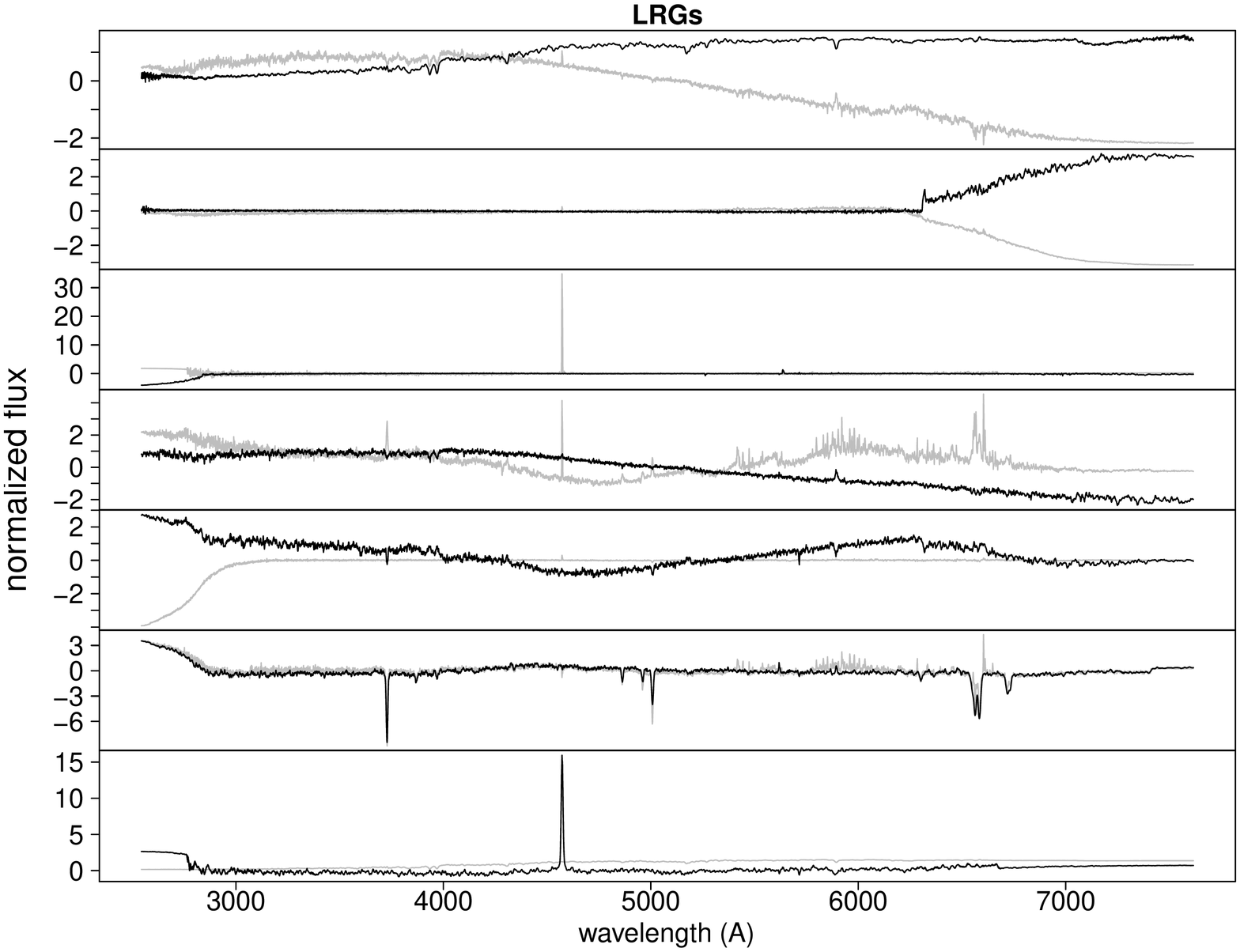}
}
\end{figure}

\clearpage

\clearpage

\clearpage

\clearpage

\clearpage

\clearpage

\clearpage

\clearpage

\clearpage

\end{document}